\documentclass[12pt, a4paper]{article}

\usepackage[utf8]{inputenc}
\usepackage[T1]{fontenc}
\usepackage{geometry}
\usepackage{amsmath, amssymb}
\usepackage{graphicx}
\usepackage{booktabs}
\usepackage{lmodern} 
\usepackage{hyperref}
\hypersetup{
    colorlinks=true,
    linkcolor=magenta,
    filecolor=green,
    urlcolor=cyan,
    citecolor=blue,
    pdftitle={Your Title},
    pdfpagemode=FullScreen,
}

\usepackage{setspace}
\usepackage{lineno} 
\usepackage{abstract}
\usepackage{spverbatim}
\usepackage{subcaption}
\usepackage{float}
\usepackage{rotating}
\usepackage{pdflscape}  
\usepackage{xcolor}
\usepackage{tcolorbox}
\usepackage{listings}
\usepackage{xcolor} 
\usepackage{natbib}
\usepackage{microtype}

\title{\bfseries Autonomous Market Intelligence: \\
Agentic AI Nowcasting Predicts Stock Returns\thanks{
    This draft is preliminary.
    We thank James Dow, Vikas Agarwal, Juhani Linnainmaa, and Wei Jiang for helpful comments.
    Results are continuously updated at \url{https://github.com/mapledust0/AI-Stock-Nowcasting/}.
    E-mails: Zefeng Chen (\href{mailto:zefengchen@gsm.pku.edu.cn}{zefengchen@gsm.pku.edu.cn}) and Darcy Pu (\href{mailto:darcypu@pku.edu.cn}{darcypu@pku.edu.cn}).
    Errors are our own.
    }
}

\author{
    \textbf{Zefeng Chen} \quad \textbf{Darcy Pu} \\
    \textit{Guanghua School of Management, Peking University}
}

\date{Jan 2026}

\begin{document}

\maketitle

\vspace{-1em}

\begin{abstract}
\noindent
\small
Can fully agentic AI nowcast stock returns? We deploy a state-of-the-art Large Language Model to evaluate the attractiveness of each Russell 1000 stock daily, starting from April 2025 when AI web interfaces enabled real-time search. Our data contribution is unique along three dimensions. First, the nowcasting framework is completely out-of-sample and free of look-ahead bias by construction: predictions are collected at the current edge of time, ensuring the AI has no knowledge of future outcomes. Second, this temporal design is irreproducible---once the information environment passes, it can never be recreated. Third, our framework is 100\% agentic: we do not feed the model news, disclosures, or curated text; it autonomously searches the web, filters sources, and synthesises information into quantitative predictions. We find that AI possesses genuine stock selection ability, but only for identifying top winners. Longing the 20 highest-ranked stocks generates a daily Fama-French five-factor plus momentum alpha of 18.4 basis points and an annualised Sharpe ratio of 2.43. Critically, these returns derive from an implementable strategy trading highly liquid Russell 1000 constituents, with transaction costs representing less than 10\% of gross alpha. However, this predictability is highly concentrated: expanding beyond the top tier rapidly dilutes alpha, and bottom-ranked stocks exhibit returns statistically indistinguishable from the market. We hypothesise that this asymmetry reflects online information structure: genuinely positive news generates coherent signals, while negative news is contaminated by strategic corporate obfuscation and social media noise.
\end{abstract}


\newpage

\onehalfspacing

\section{Introduction}

Can fully agentic AI synthesise complex, real-time information to generate actionable insights? We address this question by exploiting financial markets as a uniquely demanding laboratory. The ``efficient markets hypothesis'' posits that prices rapidly incorporate all available information, making the market an adversarial environment where systematic outperformance is extraordinarily difficult. In this context, we deploy a state-of-the-art LLM as an autonomous investment analyst. 
Starting in April 2025, on each trading day, we task the model with evaluating the entire Russell 1000 universe.
For each stock, it autonomously searches the live web, synthesises disparate information sources, and produces a quantitative attractiveness score between -5 and 5 to rank investment opportunities.
If the AI generates profitable trading signals in this context, amid competition from sophisticated market participants, it would serve as strong evidence of its capacity for advanced information evaluation.

\begin{figure}[!ht]
    \centering
    \includegraphics[width=0.85\textwidth]{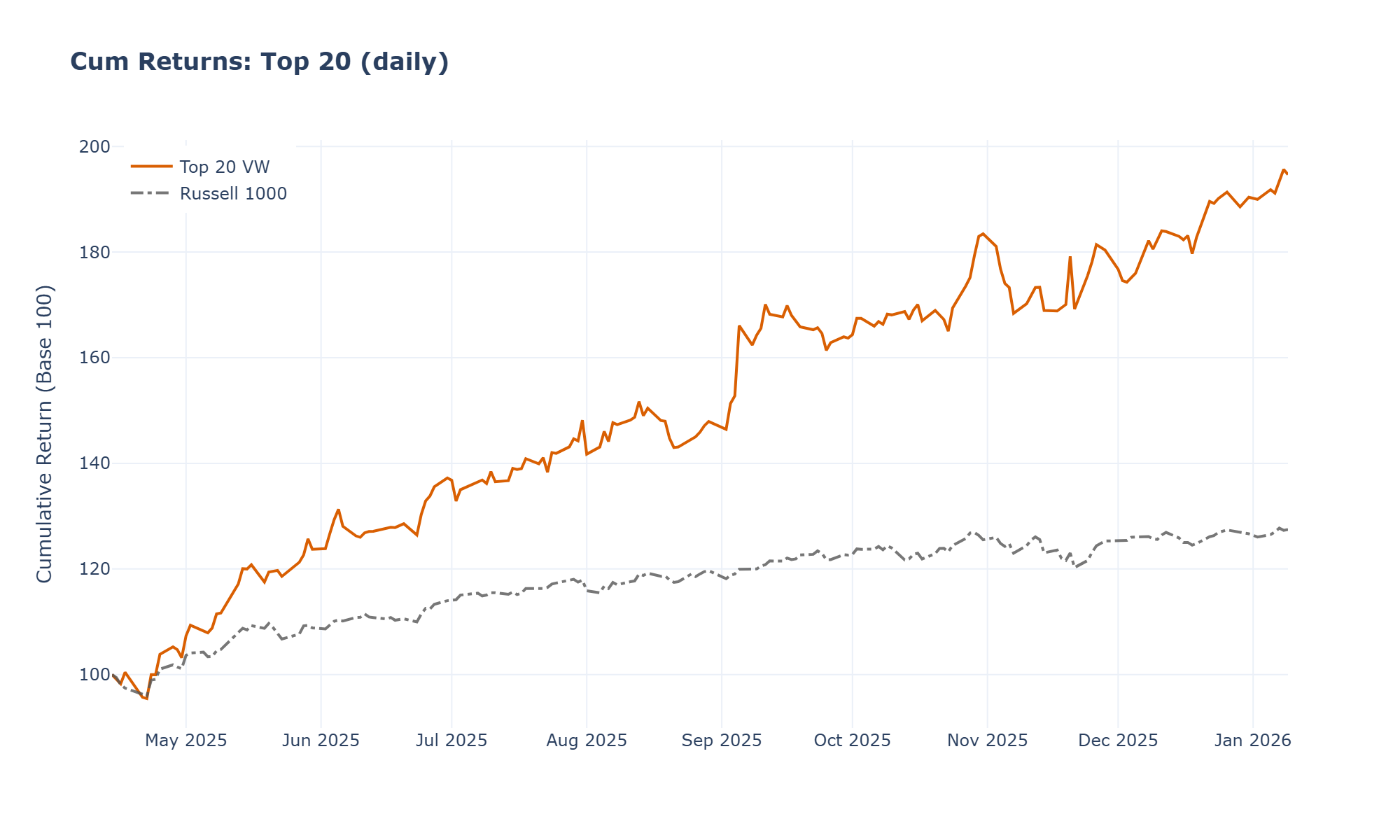}
    \caption{\textbf{Cumulative Returns of AI-Selected Portfolio vs.\ Benchmark.} This figure plots the cumulative value-weighted returns (base 100) of the Top-20 stocks selected daily by the AI model's attractiveness score (solid orange line) against the Russell 1000 index (dashed gray line). The AI-selected portfolio accumulates approximately 50\% returns over the sample period compared to 26\% for the benchmark. The performance wedge widens consistently over time, suggesting a persistent information advantage rather than exposure to isolated extreme events. Sample period: April 2025 through January 2026.}
    \label{fig:cumret_intro}
\end{figure}

Our central finding is striking: the AI demonstrates real predictive ability, but with a pronounced asymmetry. The model excels at identifying future top performers but cannot reliably distinguish future losers from average stocks. This is clearly demonstrated in Figure~\ref{fig:cumret_intro}, which shows the cumulative returns of the Top-20 AI-selected stocks compared to the Russell 1000 benchmark. A value-weighted portfolio of the 20 stocks ranked most attractive by the AI (``Top-20 VW'') accumulates approximately 50\% returns over our nine-month sample period (April 2025--January 2026), compared to roughly 26\% for the Russell 1000 benchmark---an outperformance of nearly 24 percentage points. Crucially, these returns are strictly implementable: positions are entered at the opening auction of day $t$ using signals generated overnight, and liquidated at the opening auction of day $t+1$, ensuring that the strategy accounts for execution timing and contains no look-ahead bias. This is not merely a raw return differential: controlling for the Fama-French five factors augmented with momentum, the Top-20 portfolio generates a daily alpha of 0.184\% ($t = 2.46$), implying annualised risk-adjusted abnormal returns of approximately 46\%. The portfolio achieves an annualised Sharpe ratio of 2.43, far exceeding typical hedge fund performance and standard factor strategies.

The daily frequency of our alpha is particularly noteworthy and distinct from the standard asset pricing literature.
The vast majority of documented anomalies operate at monthly or quarterly rebalancing frequencies, a design choice necessitated by the ``iron law'' of active management: high turnover typically erodes returns through transaction costs.
\citet{novymarx2016anomalies} demonstrate that strategies requiring frequent trading often generate negative net alphas once implementation costs are considered, effectively confining most anomalies to lower-frequency horizons.
Furthermore, at the daily horizon, price movements are dominated by noise or widely available news that is instantaneously incorporated by high-frequency arbitrageurs \citep{fama1991efficient, ohara2015high}, making persistent daily predictability exceptionally rare.
Against this backdrop, our finding of a robust daily alpha---one that survives strict risk adjustment and is an order of magnitude larger than the bid-ask spreads of the underlying assets---suggests that agentic AI extracts a category of rapid, high-value information that is structurally invisible to standard low-frequency factors and too complex for traditional high-frequency algorithms.

This study makes contributions that extend beyond finance and economics to the broader scientific understanding of artificial intelligence capabilities.

\paragraph{A Unique and Non-Recreatable Dataset.} We introduce the first comprehensive dataset of agentic AI stock predictions covering the entire Russell 1000 universe---approximately 1,000 of the largest U.S.\ publicly traded companies representing 93\% of domestic equity market capitalization. Unlike any prior study, our data captures the AI's \textit{autonomous} workflow: the model decides what information to retrieve, which sources to consult, and how to synthesise heterogeneous signals into quantitative predictions. This methodological distinction is fundamental. When researchers pre-select news articles or firm disclosures for AI analysis, resulting predictability confounds the model's analytical capacity with the researcher's information curation. Our agentic framework eliminates this attribution problem entirely.

Crucially, this dataset can never be re-created. The nowcasting paradigm we implement---querying the AI at the current edge of time with live web search enabled---captures a unique moment in the co-evolution of AI capabilities and financial markets. Once this temporal window closes, the opportunity to observe how contemporary AI systems process real-time market information is lost forever. Future researchers cannot go back and ask today's AI models to make predictions about today's stocks with today's information. This irreproducibility lends urgency to documenting these findings while the experimental conditions exist.

\paragraph{Financial Markets as a Natural Laboratory for AI Evaluation.} We do not study AI stock return predictability merely because profitable trading strategies are intrinsically interesting. Rather, we exploit financial markets as a uniquely demanding testbed for evaluating AI information-processing capabilities. Several features make equity markets ideal for this purpose. First, outcomes are unambiguous and precisely measurable: either the stock goes up or it goes down. Second, the environment is highly adversarial: prices reflect the aggregated beliefs of millions of participants, many of whom are sophisticated institutional investors with substantial resources. Third, information is noisy and strategically manipulated: corporate managers spin negative developments, social media influencers promote stocks for personal gain, and genuine signals are buried in avalanches of irrelevant content. If an AI system can extract actionable insights in this environment, it demonstrates capabilities that generalise far beyond finance---to any domain requiring synthesis of complex, contested, real-time information.

\paragraph{Addressing Look-Ahead Bias in LLM Evaluation.} A fundamental methodological challenge in evaluating LLMs is ``data contamination'': because these models are trained on vast corpora of historical internet content, they may possess implicit knowledge of the events they are asked to predict \citep{sainz2023nlp, golchin2023time}. This concern is particularly acute in finance, where major corporate events generate extensive online discussion that enters training data. We circumvent this problem through rigorous nowcasting: predictions are collected after market close and before the next trading session, using the latest available model version with enabled web search. The AI can access only information that was available to market participants at the moment of prediction, ensuring our framework is both econometrically valid and practically implementable.

\paragraph{Documenting Asymmetric AI Capabilities.} Our most intriguing finding is the pronounced asymmetry in the AI's predictive power. The model generates statistically significantly positive alphas for top-ranked stocks but cannot identify underperformers: bottom-ranked portfolios exhibit alphas statistically indistinguishable from zero across all specifications. This asymmetry appears robust---it holds for daily, weekly, monthly, and quarterly signal horizons and survives comprehensive risk adjustments.

We propose a mechanism rooted in the structure of online information. Genuinely positive news tends to be unambiguous: strong earnings, major contract wins, and successful product launches generate coherent signals across sources. Genuinely negative news, by contrast, coexists with strategic obfuscation. Corporate communications emphasise silver linings even amid deteriorating fundamentals. Retail investors on social media platforms post contrarian ``buy the dip'' encouragements. Speculative content proliferates precisely when prospects are most uncertain. The AI, trained to aggregate and synthesise information, may excel at recognising consensus-positive signals while struggling to distinguish genuine distress from noise-laden pessimism. This asymmetry has implications beyond finance: it suggests that current AI systems may be systematically better at identifying opportunities than threats, a finding relevant to applications ranging from medical diagnosis to threat detection.

The remainder of the paper proceeds as follows.
Section~\ref{sec:literature} provides an extensive review of related literature spanning machine learning in asset pricing, natural language processing in finance, and the computer science literature on AI trading agents.
Section~\ref{sec:data} describes our data collection methodology, including the agentic workflow and prompt design.
Section~\ref{sec:preview} presents the summary statistics and variable correlations.
Section~\ref{sec:results} reports our main empirical findings on the performance of the top-ranked portfolios and factor exposures.
Section~\ref{sec:cross_section} investigates the asymmetric predictability pattern across the entire cross-section.
Finally, Section~\ref{sec:conclusion} concludes with a discussion of limitations, implications for AI development, and directions for future research.

\section{Related Literature}
\label{sec:literature}

Our study bridges the gap between empirical asset pricing and artificial intelligence. We situate our contribution within three related literatures: the application of machine learning to return prediction, the evolution of textual analysis from sentiment dictionaries to large language models, and the development of autonomous agentic AI systems.

\subsection{Machine Learning in Asset Pricing}

The application of machine learning (ML) to asset pricing has developed primarily to navigate the ``factor zoo'' and address empirical challenges to the Efficient Markets Hypothesis. We organise this literature around three themes: factor model estimation, return prediction, and the role of feature engineering.

\paragraph{Factor Model Estimation.}
Foundational work by \citet{kelly2019characteristics} introduced Instrumented Principal Component Analysis (IPCA), which allows latent risk factors to depend on observable firm characteristics while maintaining factor model discipline. This framework provides a principled method for determining which characteristics proxy for exposure to systematic risks. \citet{lettau2020factors} extend this approach by incorporating a penalty term for pricing errors (RP-PCA), allowing for the estimation of time-varying factor risk premia even when the cross-section is large.

\paragraph{Return Prediction.}
A parallel literature focuses on direct return prediction using flexible functional forms. \citet{gu2020empirical} provide a comprehensive comparison of ML methods, finding that neural networks and gradient-boosted trees substantially outperform linear models by capturing complex nonlinear interactions among predictors. Recent studies have further refined these predictive technologies: \citet{chen2024deep} apply deep neural networks with adversarial training objectives to improve out-of-sample stability, while \citet{bryzgalova2025forest} develop tree-based methods that automatically discover characteristic interactions. Innovative approaches have also expanded the input space beyond tabular data; for instance, \citet{jiang2023reimagining} apply convolutional neural networks to stock price charts to extract visual patterns without pre-specified technical indicators.

\paragraph{Feature Engineering and Information Sets.}
Despite these methodological advances, a critical insight from recent work is that the choice of input variables---``feature engineering''---remains the dominant driver of performance. \citet{li2025machine} demonstrate that carefully constructed interaction terms often matter more than the complexity of the algorithm itself, while \citet{avramov2023machine} argue that economic restrictions are necessary to discipline ML predictions. Our study departs from this literature by addressing a shared limitation: in all prior studies, the researcher pre-selects the information set. Whether inputs are firm fundamentals or chart images, humans decide what the algorithm sees. In contrast, our agentic framework delegates the information selection process to the AI itself, allowing the model to autonomously search, filter, and synthesise information from the live web.

\subsection{Textual Analysis and Large Language Models in Finance}

The literature on extracting value from unstructured financial text has progressed through two distinct phases: traditional sentiment analysis and modern return prediction using Large Language Models (LLMs).

\paragraph{Sentiment Analysis and Attention Measures}
Early research focused on \textbf{sentiment analysis} using dictionary-based methods. Tetlock \cite{tetlock2007giving} established that negative media content predicts downward pressure on market prices, while Loughran and McDonald \cite{loughran2011liability} demonstrated that finance-specific dictionaries are necessary to capture the nuances of corporate disclosures. Subsequent studies applied supervised machine learning to refine these signals, moving beyond simple sentiment to measure \textbf{investor attention} and narrative structure. For instance, Manela and Moreira \cite{manela2017news} constructed a news-implied volatility index (NVIX) that predicts returns by capturing disaster concerns, and Bybee et al. \cite{bybee2023narrative, bybee2024business} developed indices tracking economic narratives, showing that the structure of news coverage contains predictive content distinct from raw sentiment.

\paragraph{Return Prediction via Large Language Models}
The advent of LLMs has shifted the focus from measuring sentiment to directly \textbf{predicting returns}. Unlike dictionary methods that rely on pre-specified word lists, LLMs can interpret semantic context and nuance. Lopez-Lira and Tang \cite{lopez2023can} demonstrate that GPT-based models can forecast stock price movements by analyzing news headlines, significantly outperforming traditional sentiment scores. Similarly, Chen et al. \cite{chen2022expected} show that deep learning embeddings of corporate disclosures and analyst reports capture return-relevant information that standard NLP methods miss, particularly within the first principal components of the embedding space. Beyond direct return prediction, LLMs have also shown promise in fundamental analysis, such as deciphering managerial tone in earnings calls \cite{kim2024bloated} and performing comprehensive financial statement analysis \cite{kim2024financial}.

\paragraph{Methodological Challenges: Strategic Disclosure and Look-Ahead Bias}
Despite these advances, a fundamental challenge spans this literature: the adversarial nature of the information environment. Corporate managers strategically obfuscate negative news \cite{verrecchia1983discretionary, li2008annual}, creating a ``readability gap'' that simple models may fail to penetrate. More critically, the validity of LLM-based return prediction is frequently compromised by \textbf{data contamination}. Because these models are trained on vast corpora of historical internet data, they may possess implicit knowledge of the specific outcomes they are asked to predict \cite{sainz2023nlp, golchin2023time}. Our study addresses this by implementing a strict ``nowcasting'' design, querying the AI at the current edge of time to ensure predictions rely solely on information available to market participants at the moment of decision.

\subsection{High-Frequency Return Predictability and Market Efficiency}

A substantial literature documents the difficulty of generating abnormal returns at high frequencies. \citet{chordia2014have} provide comprehensive evidence that capital market anomalies have attenuated dramatically in the recent era of high liquidity and trading activity, with daily anomaly profits essentially vanishing since the early 2000s. \citet{mclean2016does} show that academic publication of anomalies leads to substantial post-publication decay in returns, as sophisticated investors trade against documented patterns. \citet{harvey2016and} raise the bar for statistical significance in asset pricing, arguing that many published anomalies reflect data mining rather than genuine predictability.

The rise of algorithmic trading has further compressed short-horizon profit opportunities. \citet{boehmer2021algorithmic} document that algorithmic trading improves market quality and eliminates short-lived mispricings across 42 equity markets worldwide. \citet{hendershott2011does} show that algorithmic trading narrows spreads and reduces adverse selection, making it harder for any single participant to extract informational rents.

Our finding of persistent daily alpha in 2025--2026 thus represents a notable departure from these well-documented patterns. The fact that an agentic AI can generate statistically significant abnormal returns at daily frequency---after decades of efficiency improvements, the proliferation of quantitative strategies, and the widespread adoption of algorithmic trading---suggests that these systems possess information-processing capabilities qualitatively different from prior technologies.

\subsection{Agentic AI and Trading Systems in Computer Science}

A burgeoning computer science literature develops autonomous trading agents powered by LLMs, though this work often suffers from methodological limitations we aim to address.

\paragraph{Multi-Agent Trading Frameworks.} Recent studies have introduced frameworks where LLM agents simulate trading desks and market makers. \cite{xiao2024tradingagents} propose ``TradingAgents,'' and \cite{cheng2024empirical} develop a similar system for asset pricing. While these studies show that agent communication can surface return-relevant information, they are typically constrained by small sample sizes (often fewer than 50 stocks) and short evaluation periods, lacking the rigorous out-of-sample statistical inference standard in financial economics.

\paragraph{LLM-Based Trading Bots.} Numerous papers propose specific LLM-powered trading systems, including FinGPT \citep{yang2023fingpt}, BloombergGPT \citep{wu2023bloomberggpt}, and others \citep{zhang2023instruct, li2023tradinggpt, wang2024finagent, koa2024learning}. While systems like ``FinMem'' \citep{yu2023finmem} and ``LLM-Factor'' \citep{deng2023llmfactor} demonstrate creative engineering, they are frequently evaluated on negligible samples (e.g., 5--20 stocks) over brief windows, precluding robust scientific conclusions about AI trading capabilities.

\paragraph{Our Contribution.} Our study fundamentally departs from this literature in scale and design. Unlike studies limited to a handful of stocks, we evaluate AI predictions across the entire Russell 1000 universe ($\approx$ 155,000 predictions by Jan 2026) with rigorous risk adjustment and Newey-West statistical inference. Furthermore, distinctly from systems that process pre-selected data, our agentic framework autonomously searches the live web, isolating the AI's genuine information-discovery capabilities from researcher curation.

\subsection{AI Capabilities and Market Evaluation}

Our findings bridge the gap between computer science theories of AI capability and empirical asset pricing, contributing a novel economic framework for evaluating agentic intelligence.

\paragraph{Emergent Capabilities in Finance.} \cite{wei2022emergent} and \cite{bubeck2023sparks} document that advanced reasoning is an ``emergent'' property of large-scale models. We contribute to this literature by quantifying the \textit{economic value} of these emergent behaviors. We show that the ability to autonomously search, filter, and synthesise heterogeneous market signals is not explicitly trained but emerges as a financially valuable capability, generating alpha that persists even after controlling for standard risk factors.

\paragraph{Markets as Adversarial Benchmarks.} While standard AI benchmarks suffer from saturation and data contamination \citep{sainz2023nlp, hendrycks2021measuring}, financial markets provide a dynamic, adversarial laboratory. By deploying agents in a strict nowcasting framework, we establish a ``living benchmark'' that avoids the ``time travel'' issues prevalent in static evaluations \citep{golchin2023time}. Our dataset captures a unique, irreproducible snapshot of AI capabilities interacting with real-time market efficiency.

\paragraph{Theoretical Foundations and Asymmetric Limits.} \cite{kelly2025artificial} provide a theoretical basis for AI in asset pricing, suggesting models can recover latent factor structures. Our empirical contribution validates this potential but reveals a critical structural limitation: a pronounced asymmetry where AI identifies opportunities (winners) far better than threats (losers). This finding refines the theoretical understanding of AI in finance, suggesting that current architectures function optimally as information filters for capital allocation rather than as symmetric arbitrageurs.

\newpage
\newpage
\section{Data Contribution}
\label{sec:data}

\subsection{Data Collection Philosophy: The ``Agentic'' Workflow}

Our data collection methodology is built upon a fully ``agentic'' workflow that leverages the real-time capabilities of state-of-the-art Large Language Models (LLMs).

In contrast to passive studies or those suffering from look-ahead bias, our study employs a fully \textbf{``agentic'' workflow} using the web interface of a leading US AI model (we always use the latest version of this model).
We do not pre-select news or feed text to the model.
Our process is as follows:

\begin{itemize}
    \item \textbf{Interface Interaction:} For each stock in our universe, we access the web interface of a leading US-based AI model and input a standardised, expert-level prompt.
    \item \textbf{Enabled Search Function:} We explicitly enable the model's online search capability.
    \item \textbf{Autonomous Execution:} Once prompted, the model itself autonomously generates search queries, browses the live internet (including X/Twitter and financial news sites), filters for relevant events, and synthesises this information into quantitative predictions.
\end{itemize}

This distinction is critical: the AI is not just analyzing text we provide; it is deciding \textit{what} to read and \textit{how} to interpret it in real-time.
Consequently, our live, agentic nowcasting framework eliminates both of the aforementioned biases:

\begin{itemize}
    \item \textbf{Eliminating Attribution Bias:} By enabling the AI model's web interface to perform its own search and filtering, we remove the researcher from the information selection loop.
    Any predictive signal is therefore 100\% attributable to the AI model's autonomous processing, not human curation.
    \item \textbf{Eliminating Look-Ahead Bias:} We perform true out-of-sample nowcasting.
    We utilize the latest model to predict returns at the current edge of time (daily), ensuring the model has absolutely no knowledge of the future outcomes it is predicting.
\end{itemize}

\subsection{Stock Universe and Frequency}
We focus on the \textbf{Russell 1000} index, which accounts for approximately 93\% of the U.S. equity market capitalization.
This ensures our results are applicable to large, liquid stocks and are not driven by micro-cap anomalies.
We collect data at a daily frequency.
To guarantee that the AI signals produce a genuinely implementable trading strategy, we adhere to a strict temporal protocol.
All inputs are entered after the market close (4:00 PM ET) on day $t-1$, and data collection is completed before the market open on day $t$.
This design confirms that all signals are available prior to the opening auction, eliminating look-ahead bias and validating execution feasibility.
The sample covers the period from April 2025 through January 2026.

\subsection{Prompt and Variable Construction}

For every stock in the Russell 1000 index, in each day prior to market open, we input an identical prompt into the web interface to query four distinct dimensions of market analysis:

\begin{itemize}
    \item \textbf{The Attractiveness Score (The Primary Signal):} The AI model is asked to rate the attractiveness of purchasing the stock across multiple signal horizons on a scale from $-5$ (Strong Sell) to $+5$ (Strong Buy), relative to all US stocks in the entire history.
    We collect scores for signal horizons of 1 day, 1 week, 1 month, 3 months, 6 months, and 1 year.
    This is our key variable for empirical testing, especially the portfolio sorting analyses.

    \item \textbf{Price and Fundamental Forecasts:} Unlike simple sentiment scores, we require the model to output precise numerical predictions.
    These include predicted closing prices for signal horizons ranging from today to one year out, Earnings Per Share (EPS) predictions for the next 5 fiscal years, and an earnings report rating ($-5$ to $+5$) predicting whether the next report will beat or miss analyst expectations.

    \item \textbf{Market Sentiment and Divergence:} The model is explicitly instructed to analyze social sentiment, particularly from X (formerly Twitter).
    It outputs a \textbf{Sentiment Score} ($-5$ to $+5$) reflecting the overall market mood and a \textbf{Divergence Score} ($-5$ to $+5$) measuring investor disagreement.
    A low divergence score implies high volatility or confusion, while a high score implies consensus.

    \item \textbf{Benchmark Control:} To control for the AI's evaluation of general market conditions, our prompt asks the model to provide an attractiveness rating for the Russell 1000 index itself (a market proxy) for the same signal horizons.
    This allows us to normalise single-stock scores against the model's view of the broader market and eliminate the possibility that the AI might view the world a bit more or less optimistically from time to time randomly.
\end{itemize}

\subsection{Data Extraction and Reliability}

The model is instructed to summarise its analysis into a Python-readable list format at the end of every response.
We use Regular Expressions (Regex) to extract this 40-item list automatically from the chat logs.
This allows us to convert the model's unstructured web analysis into a structured panel dataset.

To test whether the AI model produces random forecasts, we repeat the querying process multiple times for Dow Jones constituent stocks (which account for 30\% of total US market capitalization) on randomly selected days.
Finally, the daily financial data for the Russell 1000 stocks are obtained from LSEG DataStream.

\section{Data Preview}
\label{sec:preview}
\subsection{Summary Statistics}
Table \ref{tab:summary_stats} presents summary statistics for the key variables used in our analysis.
The sample comprises nearly 156,000 daily stock-level observations spanning April 2025 through January 2026, covering all constituents of the Russell 1000 index.

\paragraph{AI-Generated Signals.}
Panels A through C report the outputs generated by the AI model in response to each daily nowcasting prompt.
As detailed in Appendix \ref{appendix_prompt}, signals are collected overnight before market open.

Panel A reports statistics for our primary independent variables: individual stock attractiveness scores bounded between $-5$ (Strong Sell) and $+5$ (Strong Buy).
While mean attractiveness hovers near zero across all signal horizons, the distribution reveals a pronounced ``term structure'' in the model's optimism.
Median scores increase monotonically from $0.50$ at the 1-day signal horizon to $2.20$ at the 1-year signal horizon, suggesting the AI systematically expresses greater conviction in longer-term equity appreciation---consistent with the historical equity risk premium.
Standard deviations range from 1.09 to 1.29, indicating substantial cross-sectional dispersion; the model does not simply revert to a neutral consensus but actively differentiates among stocks.

Panel B presents the corresponding attractiveness scores for the Russell 1000 index, which serve as our market benchmark control.
The markedly lower volatility of index-level scores (SD $\approx 0.64$--$0.77$ versus $\approx 1.09$--$1.29$ for individual stocks) confirms that the model distinguishes between idiosyncratic stock characteristics and broad market conditions---a prerequisite for meaningful stock selection signals.

Panel C summarises the model's assessment of aggregate market sentiment.
The \textit{market\_sentiment} score has a mean of 1.35 and substantial variation of 1.59, reflecting the model's dynamic view of the macro environment.
The \textit{market\_divergence} variable tracks the degree of disagreement between the specific stock's outlook and the broader market mood.

\paragraph{Financial Data.}
Panel D reports the open-to-open return---our primary dependent variable in subsequent portfolio sorting analyses---along with market characteristics that describe the cross-sectional composition of our sample.
The \textbf{open-to-open return}, computed as the percentage return from day $t$'s market open to day $t+1$'s market open, serves as the outcome variable against which we evaluate the predictive content of AI-generated signals.
This return window is dictated by our data collection protocol: AI attractiveness ratings are generated overnight, after day $t-1$'s market close (4:00 PM ET) and before day $t$'s market open.
Consequently, using open-to-open returns ensures strict temporal separation between signal observation and return realization, eliminating any possibility of look-ahead bias.

The sample focuses on a highly liquid universe: median market capitalization is approximately \$15.2 billion and median daily dollar trading volume is \$365 million.
Bid-ask spreads are extremely tight, with a median of just 3.41 basis points, ensuring our results are not confounded by microstructure effects or micro-cap anomalies.
The open-to-open returns exhibit the expected distributional properties, with a mean near zero, slight positive skewness (1.55), and fat tails consistent with typical equity return distributions.

\begin{landscape}
\begin{table}[!ht]
    \centering
    \resizebox{\textwidth}{!}{
\begin{tabular}{l c c c c c c c c c c c}
\hline\hline
Variable & Count & Min & 1\% & 25\% & Median & 75\% & 99\% & Max & Mean & SD & Skew \\
\hline
\multicolumn{12}{l}{\textbf{Panel A: Individual Stock Attractiveness}} \\
\hline
attractiveness\_1d & 155,891 & -5.00 & -2.50 & -0.50 & 0.50 & 1.20 & 3.75 & 4.92 & 0.38 & 1.29 & -0.01 \\
attractiveness\_1w & 155,887 & -5.00 & -2.30 & -0.50 & 0.30 & 1.20 & 3.55 & 4.95 & 0.38 & 1.23 & 0.06 \\
attractiveness\_1m & 155,887 & -5.00 & -2.00 & 0.10 & 0.70 & 1.70 & 3.80 & 5.00 & 0.82 & 1.18 & -0.02 \\
attractiveness\_1q & 155,885 & -5.00 & -1.50 & 0.78 & 1.20 & 2.05 & 4.10 & 5.00 & 1.34 & 1.13 & -0.11 \\
attractiveness\_2q & 155,884 & -5.00 & -1.00 & 1.20 & 1.80 & 2.50 & 4.35 & 5.00 & 1.81 & 1.09 & -0.21 \\
attractiveness\_1y & 155,884 & -5.00 & -0.70 & 1.50 & 2.20 & 3.10 & 4.70 & 5.00 & 2.28 & 1.14 & -0.33 \\
\hline
\multicolumn{12}{l}{\textbf{Panel B: Russell 1000 Attractiveness}} \\
\hline
attractiveness\_russell\_1d & 136,561 & -4.32 & -0.50 & 0.30 & 0.50 & 0.80 & 2.50 & 5.00 & 0.62 & 0.64 & 1.00 \\
attractiveness\_russell\_1w & 136,557 & -4.30 & -0.70 & 0.20 & 0.30 & 0.78 & 2.60 & 4.60 & 0.57 & 0.70 & 1.16 \\
attractiveness\_russell\_1m & 136,554 & -4.65 & -0.50 & 0.50 & 0.70 & 1.01 & 2.90 & 4.50 & 0.88 & 0.69 & 1.09 \\
attractiveness\_russell\_1q & 136,553 & -4.88 & -0.20 & 0.80 & 1.00 & 1.50 & 3.20 & 4.70 & 1.21 & 0.70 & 1.06 \\
attractiveness\_russell\_2q & 136,553 & -4.30 & 0.15 & 1.20 & 1.30 & 1.80 & 3.60 & 4.89 & 1.54 & 0.72 & 1.00 \\
attractiveness\_russell\_1y & 136,541 & -4.50 & 0.35 & 1.50 & 1.68 & 2.20 & 4.10 & 5.00 & 1.89 & 0.77 & 0.83 \\
\hline
\multicolumn{12}{l}{\textbf{Panel C: Market Sentiment}} \\
\hline
market\_sentiment & 126,847 & -5.67 & -2.75 & 0.50 & 1.35 & 2.45 & 4.25 & 5.00 & 1.14 & 1.59 & -0.47 \\
market\_divergence & 126,845 & -4.85 & -3.40 & -1.50 & 0.23 & 2.00 & 4.20 & 5.00 & 0.38 & 2.03 & 0.20 \\
\hline
\multicolumn{12}{l}{\textbf{Panel D: Financial Metrics}} \\
\hline
Market Cap & 199,164 & 62.25 & 1,203.24 & 7,438.89 & 15,218.96 & 42,752.28 & 917,135.74 & 5,031,072.00 & 66,241.26 & 270,889.61 & 10.89 \\
Dollar Trading Volume & 190,749 & 18.45 & 2,142.40 & 86,315.66 & 364,962.36 & 1,756,075.29 & 215,014,117.11 & 14,955,206,520.00 & 13,247,706.41 & 177,392,852.98 & 35.45 \\
Spread (bps) & 193,187 & 0.01 & 0.21 & 1.58 & 3.41 & 6.37 & 25.41 & 131.58 & 4.94 & 5.60 & 4.73 \\
Open-to-Open Return & 202,462 & -0.67 & -0.07 & -0.01 & 0.00 & 0.01 & 0.08 & 1.29 & 0.00 & 0.03 & 1.55 \\
\hline
\hline
\end{tabular}
}
    \vspace{0.2cm}
    \caption{\textbf{Summary Statistics}. This table reports summary statistics for daily variables used in the analysis. 
    The sample period covers April 2025 through January 2026 for Russell 1000 constituents. 
    \textbf{Panels A--C} report AI-generated outputs from the daily nowcasting prompt (see Appendix \ref{appendix_prompt}). 
    \textit{Attractiveness Scores} (Panels A--B) are bounded $[-5, +5]$, where $-5$ indicates ``Strong Sell'' and $+5$ indicates ``Strong Buy.''
    \textit{Market Sentiment} (Panel C) reflects the model's aggregate market outlook; \textit{Market Divergence} measures the difference between stock-specific and market-level attractiveness.
    \textbf{Panel D} reports the dependent variable and market characteristics: 
    \textit{Open-to-Open Return} is the percentage return from day $t$'s open to day $t+1$'s open and serves as the primary dependent variable;
    \textit{Market Cap} is in millions of USD; 
    \textit{Dollar Trading Volume} is daily volume $\times$ price (thousands USD); 
    \textit{Spread (bps)} is the bid-ask spread relative to the mid-price in basis points.}
    \label{tab:summary_stats}
\end{table}
\end{landscape}

\subsection{Correlation Analysis}

Table \ref{tab:corr_matrix} reports the pairwise correlation coefficients for a select subset of our key variables.
Several patterns emerge that support the distinctiveness of our AI-generated signals.

First, the \textbf{Attractiveness Scores} exhibit high serial correlation across signal horizons but remain distinct from market characteristics.
The 1-day attractiveness score (\textit{Attr 1D}) is highly correlated with the 1-week (\textit{Attr 1W}, $\rho = 0.97$) and 1-month (\textit{Attr 1M}, $\rho = 0.91$) scores, indicating consistency in the model's short-to-medium term outlook.
However, the correlation drops significantly when compared to the benchmark \textit{Russell 1000 Attractiveness} score ($\rho = 0.27$), confirming that the stock-specific signal contains substantial idiosyncratic information beyond the model's general market view.

Second, the correlation between AI signals and standard market characteristics is remarkably low.
The correlation between \textit{Attr (1D)} and \textit{Market Cap} is only $0.16$, and its correlation with \textit{Dollar Trading Volume} is nearly zero ($0.06$).
This suggests that the AI's recommendations are not simply proxies for size or liquidity factors.
Similarly, the correlation with \textit{Spread} is negative ($-0.17$), indicating a slight preference for more liquid stocks, but not to an extent that would dominate the signal.

Third, the \textit{Sentiment} score is moderately correlated with the stock attractiveness scores ($\rho \approx 0.69$--$0.71$), suggesting that the model's overall market mood influences its specific stock recommendations, yet the divergence is sufficient to allow for contrarian calls.
Finally, the \textit{Probability of Beating Consensus} shows a robust positive correlation with attractiveness ($\rho \approx 0.60$), implying that the model's fundamental earnings outlook is a key driver of its buy/sell recommendations.

Importantly, the correlation between AI attractiveness scores and the open-to-open return---our dependent variable---is essentially zero in the unconditional correlation matrix.
This near-zero correlation suggests that the AI signal does not exhibit a simple linear relationship with future returns across the entire cross-section of stocks.
Whether the signal nonetheless contains economically meaningful predictive content---particularly in the tails of the distribution---is an empirical question we address in the portfolio sorting analyses that follow.

\begin{table}[!ht]
    \centering    
    \scriptsize    
    \resizebox{\textwidth}{!}{
\begin{tabular}{l c c c c c c c c c c c }
\hline\hline
Variable & (1) & (2) & (3) & (4) & (5) & (6) & (7) & (8) & (9) & (10) & (11) \\
\hline
Attr (1D) & 1.00 &  &  &  &  &  &  &  &  &  &  \\
Attr (1W) & $0.97^{***}$ & 1.00 &  &  &  &  &  &  &  &  &  \\
Attr (1M) & $0.91^{***}$ & $0.95^{***}$ & 1.00 &  &  &  &  &  &  &  &  \\
Russell Attr (1D) & $0.27^{***}$ & $0.31^{***}$ & $0.31^{***}$ & 1.00 &  &  &  &  &  &  &  \\
Sentiment & $0.69^{***}$ & $0.71^{***}$ & $0.71^{***}$ & $0.22^{***}$ & 1.00 &  &  &  &  &  &  \\
Divergence & $0.36^{***}$ & $0.37^{***}$ & $0.36^{***}$ & $0.12^{***}$ & $0.44^{***}$ & 1.00 &  &  &  &  &  \\
Prob Beat (1Q) & $0.60^{***}$ & $0.64^{***}$ & $0.67^{***}$ & $0.25^{***}$ & $0.63^{***}$ & $0.32^{***}$ & 1.00 &  &  &  &  \\
Market Cap & $0.16^{***}$ & $0.16^{***}$ & $0.16^{***}$ & $0.04^{***}$ & $0.14^{***}$ & $0.02^{***}$ & $0.12^{***}$ & 1.00 &  &  &  \\
Dollar Vol & $0.06^{***}$ & $0.06^{***}$ & $0.06^{***}$ & $0.02^{***}$ & $0.05^{***}$ & $-0.01^{***}$ & $0.05^{***}$ & $0.54^{***}$ & 1.00 &  &  \\
Spread & $-0.17^{***}$ & $-0.17^{***}$ & $-0.17^{***}$ & $0.01^{*}$ & $-0.14^{***}$ & $-0.07^{***}$ & $-0.12^{***}$ & $-0.09^{***}$ & $-0.03^{***}$ & 1.00 &  \\
Ret (Open-Open) & $0.00$ & $0.00$ & $-0.00$ & $-0.01^{***}$ & $-0.01^{**}$ & $-0.01^{**}$ & $0.01^{**}$ & $0.01^{***}$ & $0.01^{***}$ & $-0.00^{**}$ & 1.00 \\
\hline
\hline
\end{tabular}
}    
    \vspace{0.1cm}
    \caption{\textbf{Correlation Matrix of Key Variables}. This table reports Pearson correlation coefficients for a subset of key AI-generated signals and market characteristics. 
        $^{*}$ $p<0.1$; $^{**}$ $p<0.05$; $^{***}$ $p<0.01$.
        \textit{Attr (1D/1W/1M)}: Stock Attractiveness Score at 1-day, 1-week, and 1-month signal horizons.
        \textit{Russell Attr (1D)}: Attractiveness Score for the Russell 1000 Index (1-day).
        \textit{Sentiment}: AI-generated market sentiment score.
        \textit{Prob Beat (1Q)}: Probability of beating earnings consensus (1-quarter).
        \textit{Market Cap}, \textit{Dollar Vol}, and \textit{Spread}: Market characteristics describing sample composition.
        \textit{Ret (Open-Open)}: Daily open-to-open return, serving as the dependent variable in portfolio sorting analyses.}
    \label{tab:corr_matrix}
\end{table}

\subsection{Consistency of AI Scores}
\label{subsec:consistency}

A central concern when utilising Large Language Models (LLMs) for financial analysis is the stochastic nature of text generation. Unlike traditional quantitative factors, an LLM may output slightly different scores for the same input prompt upon regeneration. To validate that our \textit{attractiveness} scores represent a stable, fundamental signal rather than random generation noise, we conduct a rigorous battery of consistency tests.

We design an experiment in which 30 randomly selected stocks are queried repeatedly (approximately 350 times per stock) across overlapping dates. We then apply five statistical tests to assess the distributional distinctness, internal reliability, and rank stability of these scores. Table \ref{tab:ai_consistency} summarises the results.

First, we verify that the model distinguishes between assets using the \textbf{Kolmogorov-Smirnov Distribution Test}. This test compares the distribution of scores for each individual stock against the aggregate population distribution. The null hypothesis is that a specific stock's scores are essentially random draws from the general population. We observe a 100\% rejection rate ($p<0.05$) across all 30 tested stocks. This confirms that the AI generates distinct, asset-specific score distributions rather than simply reverting to a generic ``market'' mean.

Second, we assess whether the signal is consistent over time for specific assets using a \textbf{Monte Carlo Permutation Test}. We compute the pooled within-group variance for each (stock, date) pair and compare it against a null distribution generated by randomly shuffling scores 1,000 times. The empirical $p$-value is less than $0.001$, indicating that the low variance we observe for a given stock on a given day is highly significant. This result implies that the model's output is driven by the specific stock-date information set, not random noise.

Third, we measure the precision of the signal using \textbf{Split-Half Reliability Analysis}, a standard psychometric validation technique. By randomly splitting the repeated observations for the same stock-date into two halves, we observe a Pearson correlation of $\rho = 0.942$. This near-perfect correlation demonstrates that the ``noise'' introduced by the LLM's stochasticity is negligible compared to the strength of the fundamental signal. Similarly, the \textbf{Rank Stability Analysis} yields a Spearman correlation of $\rho = 0.902$, confirming that the model's relative ordering of stocks remains stable across repeated generations.

Finally, the \textbf{Production Alignment Validation} confirms the integrity of our main dataset. We compare the mean scores from this experimental repeated-query dataset against the single-query scores used in our main analysis ($df0$). We find a robust positive Spearman correlation of $\rho = 0.629$. While strictly lower than the internal reliability metrics—as expected when comparing a single stochastic draw to a mean of 350 draws—this result confirms that the single-query signals used in our main analysis are representative of the model's central tendency and effectively capture the underlying signal.

\begin{table}[!ht]
    \centering
    \resizebox{\textwidth}{!}{
\begin{tabular}{l l l c}
\hline \hline
\textbf{Test Methodology} & \textbf{Description} & \textbf{Metric} & \textbf{Estimate} \\ 
\hline
    Kolmogorov-Smirnov Distribution Test & Tests if individual stock score distributions are distinct from the aggregate population. & Rejection Rate & 100\% (30/30) \\
    Monte Carlo Permutation Test & Compares within-group variance of stock-date scores against 1,000 random shuffles. & Empirical $p$-value & $< 0.001$ \\
    Split-Half Reliability Analysis & Correlates means of two random halves of repeated queries for the same stock-date. & Pearson $\rho$ & 0.942 \\
    Rank Stability Analysis & Assesses the stability of stock rankings across repeated queries on the same date. & Spearman $\rho$ & 0.902 \\
    Production Alignment Validation & Correlates experimental repeated-query means with single-query production data. & Spearman $\rho$ & 0.629 \\
\hline \hline
\end{tabular}
}

    \caption{\textbf{AI Score Consistency and Reliability Tests}. 
    This table summarises the results of five statistical tests designed to validate the stability of the AI-generated attractiveness scores. 
    The analysis uses a subsample of 30 stocks queried repeatedly (approx. 350 times per stock) to assess whether the model outputs are stochastic noise or stable signals. 
    \textit{Rejection Rate} refers to the percentage of stocks for which the null hypothesis (identical distribution to population) is rejected at the 5\% level.
    \textit{Empirical p-value} is calculated from 1,000 random permutations.
    \textit{Pearson/Spearman $\rho$} report correlation coefficients for reliability and alignment tests.}
    \label{tab:ai_consistency}
\end{table}

\section{AI Predicts Top Winners}
\label{sec:results}
We construct our implementable investment portfolios as follows. At the beginning of each trading day $t$, we rank all stocks in descending order based on the AI-generated ``Attractiveness Score'' generated during the overnight session (post-close $t-1$). We define the Top-20 portfolio as the set of $N=20$ stocks with the highest AI attractiveness scores. Ties at the inclusion threshold ($N$-th rank) are resolved by prioritising the stock with the larger market capitalization to favour liquidity. The portfolio is rebalanced daily using value-weights. To strictly eliminate look-ahead bias and ensure the strategy is implementable using only information available prior to trading, we calculate performance using open-to-open returns. Specifically, positions are assumed to be entered at the opening auction of day $t$---utilising the signals generated overnight---and held until the opening auction of day $t+1$.

For signal horizons extending beyond the daily frequency (weekly, monthly, quarterly), we employ an overlapping portfolio approach to mitigate rebalancing timing risk. Specifically, for a target holding period of $K$ days, the aggregate portfolio at time $t$ consists of $K$ sub-portfolios initiated on rolling days $t, t-1, ..., t-K+1$, each weighted equally at $1/K$.

For evaluation, we regress excess returns on the Fama-French five factors augmented with momentum:
\begin{equation}
    R_{p,t} - R_{f,t} = \alpha_p + \boldsymbol{\beta}_p' \mathbf{F}_t + \epsilon_{p,t}
\end{equation}
where $R_{p,t}$ is the value-weighted return of portfolio $p$, $R_{f,t}$ is the risk-free rate, and $\mathbf{F}_t$ is the vector of risk factors $(MKT, SMB, HML, RMW, CMA, MOM)'$. The intercept $\alpha_p$ represents abnormal return unexplained by factor exposures; a significantly positive $\alpha$ indicates the AI model possesses unique predictive information beyond standard risk premia.  Standard errors are computed using Newey-West with five lags to account for heteroskedasticity and autocorrelation.

\subsection{Top-20 Portfolio's Stellar Performance}

\begin{figure}[!ht]
    \centering
    \begin{subfigure}[t]{0.48\textwidth}
        \centering
        \includegraphics[width=\linewidth]{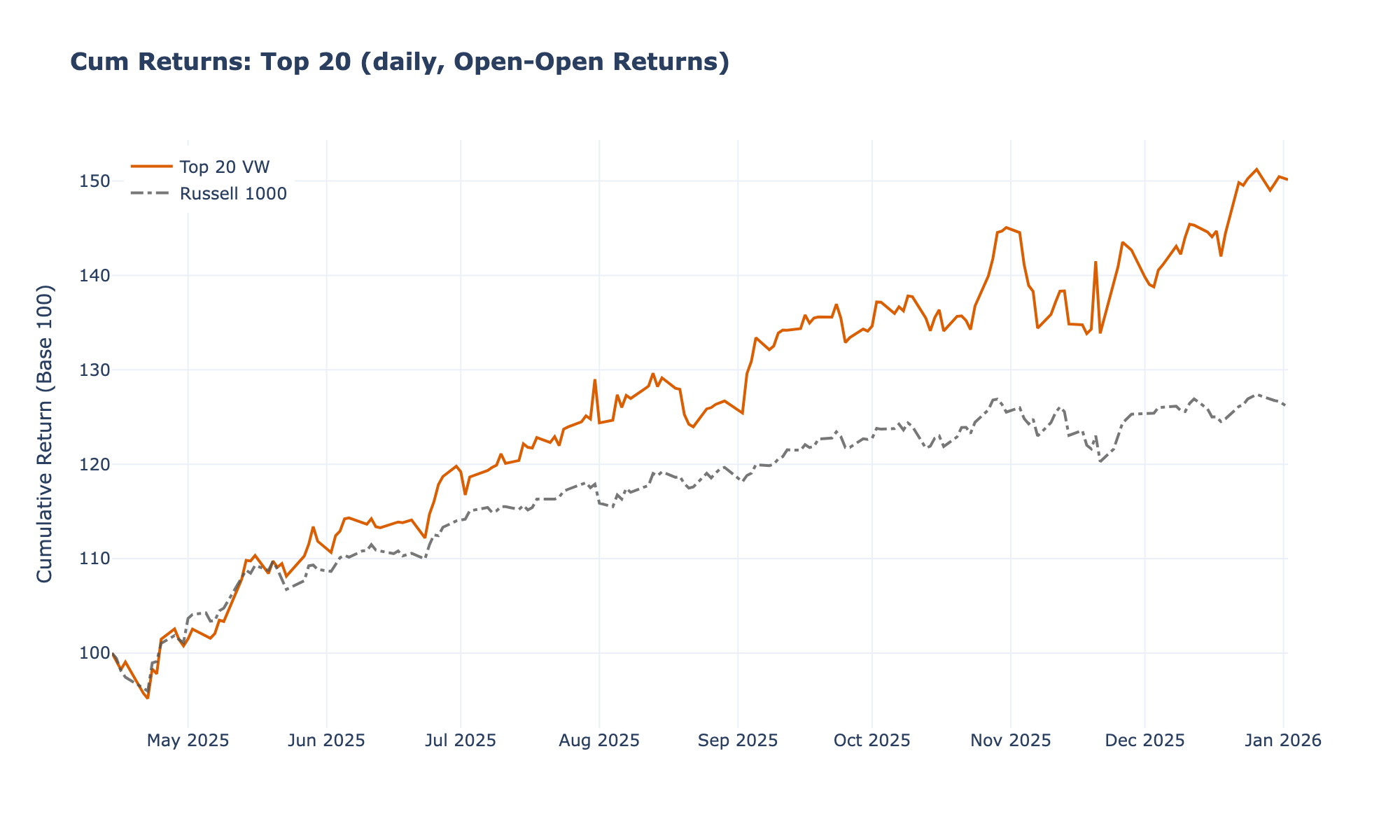}
        \caption{Daily Signal Horizon}
    \end{subfigure}
    \hfill
    \begin{subfigure}[t]{0.48\textwidth}
        \centering
        \includegraphics[width=\linewidth]{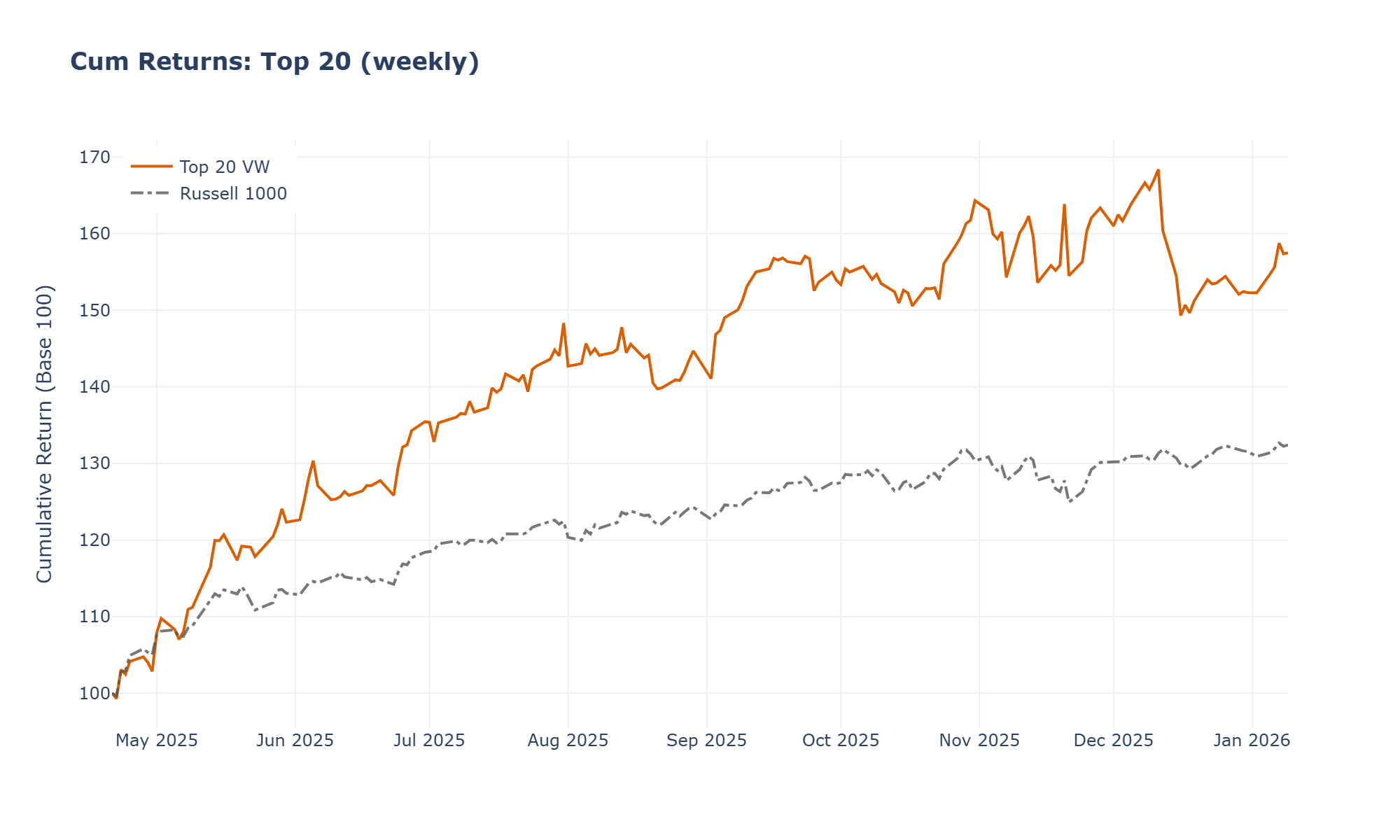}
        \caption{Weekly Signal Horizon}
    \end{subfigure}
    \vskip\baselineskip
    \begin{subfigure}[t]{0.48\textwidth}
        \centering
        \includegraphics[width=\linewidth]{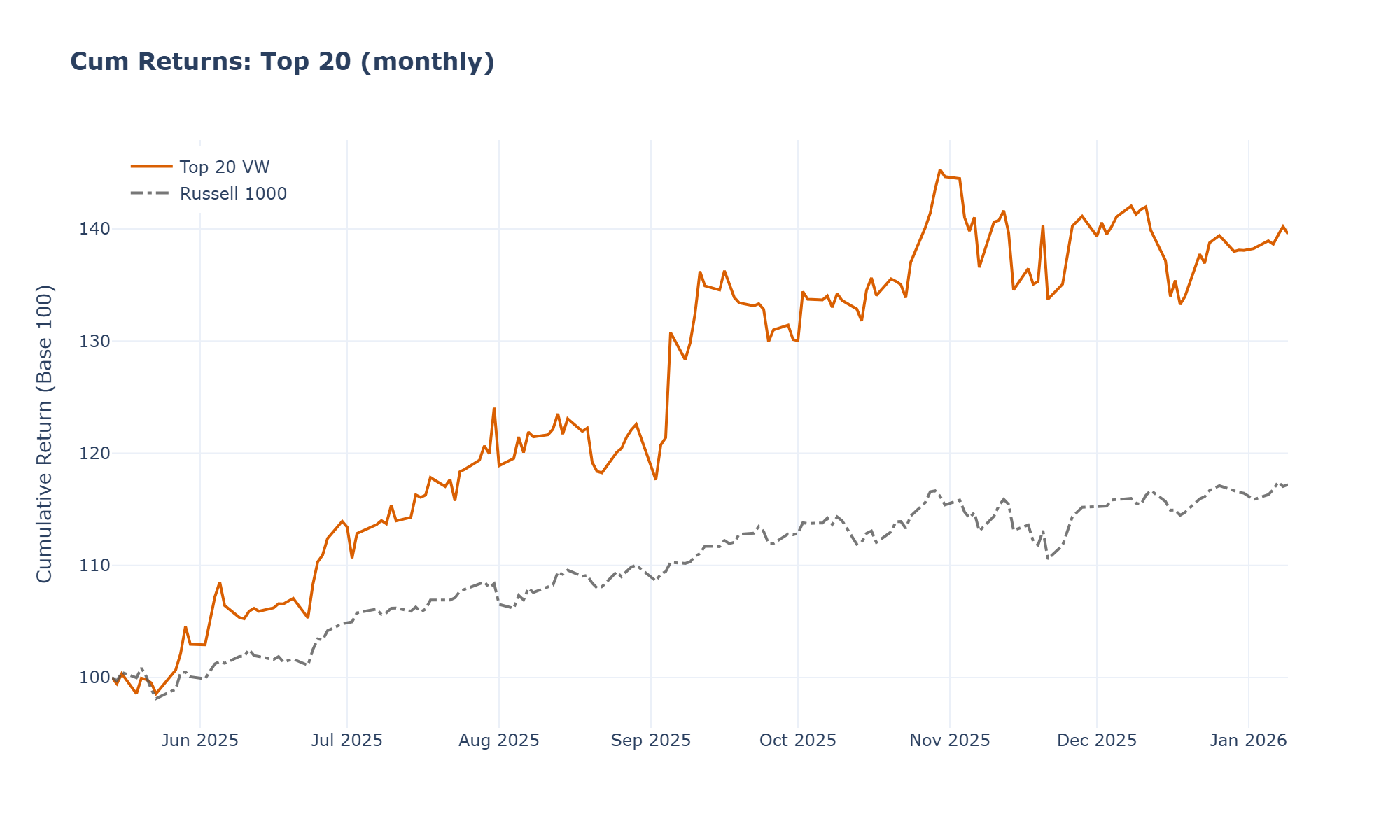}
        \caption{Monthly Signal Horizon}
    \end{subfigure}
    \hfill
    \begin{subfigure}[t]{0.48\textwidth}
        \centering
        \includegraphics[width=\linewidth]{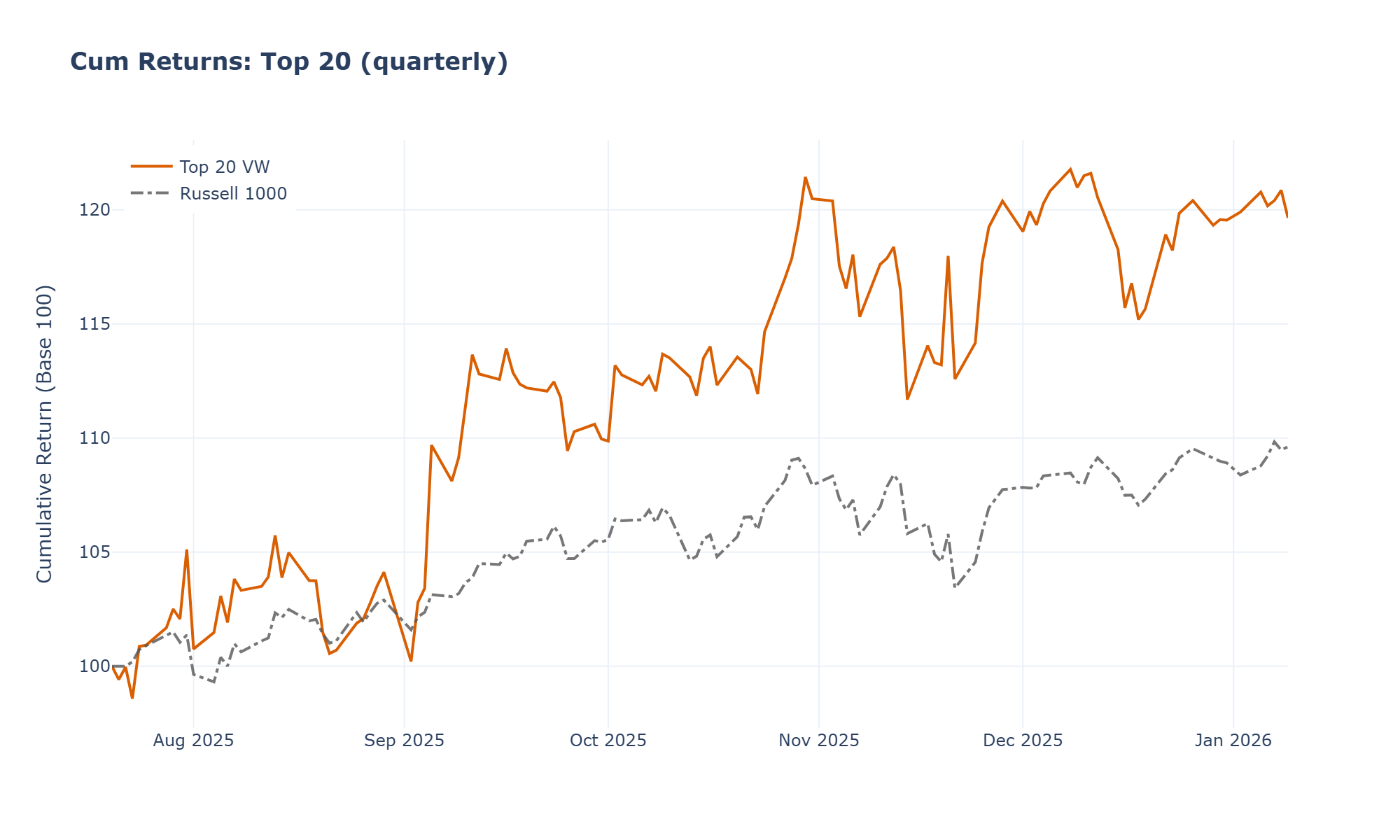}
        \caption{Quarterly Signal Horizon}
    \end{subfigure}
    \caption{\textbf{Cumulative returns of the Top-20 portfolios under different signal horizons.} This figure plots the value-weighted cumulative returns of portfolios of Top-20 stocks ranked by AI's daily, weekly, monthly, and quarterly attractiveness scores. Each panel shows the cumulative return (base 100) for the Top-20 portfolio and the Russell 1000 benchmark over the sample period.}
    \label{fig:top20_cumret_2x2}
\end{figure}

Figure \ref{fig:top20_cumret_2x2} plots the cumulative value-weighted returns of the portfolio of Top-20 stocks based on AI attractiveness rating, against the Russell 1000 benchmark. The visual evidence indicates a substantial and persistent performance wedge. Across all four signal horizons---daily, weekly, monthly, and quarterly---the AI-selected portfolio consistently decouples from the broad market, exhibiting a steady upward trajectory even during periods where the benchmark remains flat. Notably, this outperformance is not driven by a few isolated jump events but rather accumulates smoothly over the sample period, suggesting a consistent information advantage rather than exposure to high-variance tail events.

\begin{table}[!ht]
    \centering
    \begin{tabular}{llllll}
\toprule
 & (1) Alpha only & (2) CAPM & (3) FF3 & (4) FF5 & (5) FF6 \\
\midrule
$\alpha$ & 0.370*** & 0.301** & 0.319*** & 0.336*** & 0.334*** \\
 & (2.88) & (2.54) & (2.84) & (3.01) & (2.96) \\
MKT &  & 0.470*** & 0.298 & 0.340* & 0.347* \\
 &  & (2.62) & (1.46) & (1.75) & (1.82) \\
SMB &  &  & 0.078 & 0.095 & -0.013 \\
 &  &  & (0.30) & (0.35) & (-0.04) \\
HML &  &  & -0.684*** & -0.906*** & -0.883*** \\
 &  &  & (-2.94) & (-3.59) & (-3.60) \\
RMW &  &  &  & 0.245 & 0.191 \\
 &  &  &  & (0.95) & (0.70) \\
CMA &  &  &  & 0.440 & 0.237 \\
 &  &  &  & (1.10) & (0.55) \\
MOM &  &  &  &  & -0.194 \\
 &  &  &  &  & (-1.18) \\
Obs & 158 & 158 & 158 & 158 & 158 \\
\bottomrule
\end{tabular}

    \caption{\textbf{The Top-20 Portfolio (Daily Attractiveness Signal).} This table reports the results of time-series regressions of the daily value-weighted excess returns of the Top-20 portfolio (ranked by daily AI attractiveness signal) on various factor models. Each column corresponds to a different regression specification: (1) excess return regressed on a constant only; (2) CAPM (MKT); (3) Fama-French 3-factor (MKT, SMB, HML); (4) Fama-French 5-factor (MKT, SMB, HML, RMW, CMA); (5) Fama-French 6-factor (FF5 + MOM). Coefficients are reported in percent per day. Newey–West (HAC) standard errors with 5 lags are used to compute t-statistics, reported in parentheses. ***, **, and * indicate significance at the 1\%, 5\%, and 10\% levels, respectively.}
    \label{tab:top20alphas}
\end{table}

Table \ref{tab:top20alphas} formally evaluates this performance by regressing the daily excess returns of the daily-signal Top-20 portfolio on standard asset pricing factor models. The results confirm the economic and statistical significance of the AI's stock selection. In the full Fama-French six-factor specification (Column 5), the Top-20 portfolio generates a daily alpha of 0.184\% ($t = 2.46$), which is statistically significant at the 5\% level. This point estimate implies an annualised abnormal return of approximately 46\%, a magnitude that is economically large for a strategy trading strictly liquid, large-cap US equities.

The economic magnitude of these daily alphas warrants emphasis. In the extensive literature on cross-sectional return predictability, documented alphas almost universally operate at monthly or longer horizons. \citet{chordia2014have} show that most anomalies have disappeared at daily frequencies due to increased market efficiency and algorithmic trading. \citet{mclean2016does} document substantial post-publication decay in anomaly returns. Against this backdrop, generating statistically significant daily alpha of 18.4 basis points in 2025--2026 represents a notable departure from typical findings. Moreover, our strategy is fully implementable: we trade exclusively Russell 1000 constituents with median market capitalization of \$15.2 billion and median bid-ask spreads of just 1.6 basis points (see Appendix \ref{sec:trading_costs}). The gross alpha exceeds transaction costs by an order of magnitude, ensuring that our findings are not artifacts of illiquidity or microstructure frictions.

The factor loadings in Table \ref{tab:top20alphas} provide crucial insight into the ``style'' of the AI in forming this particular portfolio. First, the market beta (MKT) is remarkably low at 0.298 ($t=2.17$), indicating that the AI does not simply select high-beta stocks to amplify market returns. Instead, it constructs a defensive portfolio that is far less volatile than the market index. Second, the portfolio exhibits a strong negative loading on the value factor (HML coefficient = -0.636, $t=-4.02$), revealing a distinct preference for growth stocks. Finally, the loading on Momentum (MOM) is statistically insignificant ($-0.169$, $t=-1.33$), clarifying that the AI is not merely chasing past winners.

\subsection{Where Does AI Predict? Only the Very Top}

A natural question for any return-predictive signal is whether it identifies both winners and losers, or whether its information content is concentrated on one side.
This distinction matters for implementation: a symmetric signal supports long-short strategies, while an asymmetric signal favours long-only portfolios.
We investigate this by comparing the risk-adjusted performance of top-ranked versus bottom-ranked stocks across varying portfolio sizes.

\begin{figure}[!ht]
    \centering
    \includegraphics[width=\textwidth]{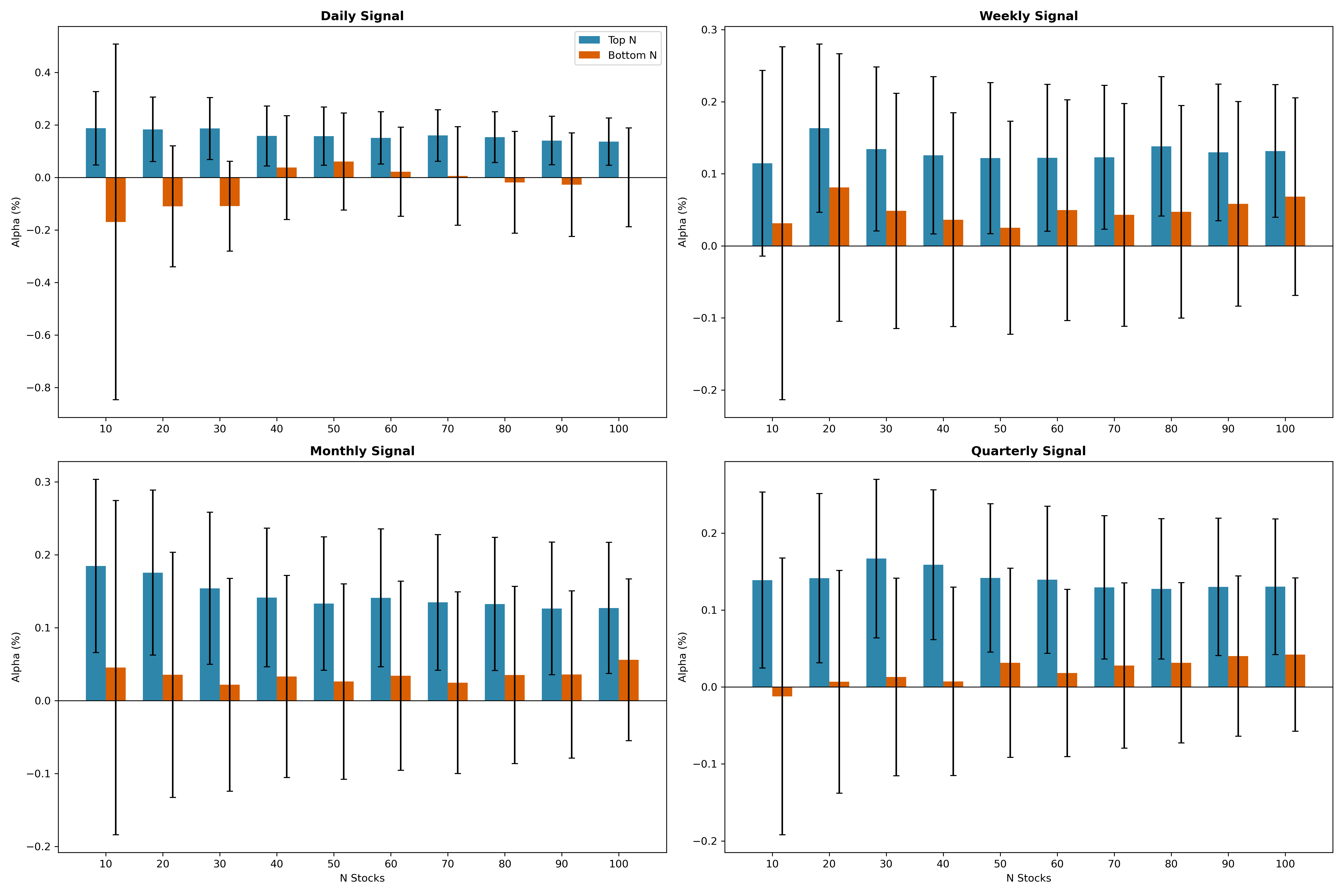}
    \caption{\textbf{Top versus Bottom Alpha Performance.} This figure reports daily Fama-French six-factor alphas (\%) for value-weighted portfolios of the top $N$ (blue) and bottom $N$ (orange) stocks ranked by AI attractiveness. Panels correspond to daily, weekly, monthly, and quarterly signal horizons. Error bars indicate 90\% confidence intervals using Newey-West standard errors (5 lags). Sample period: April 2025 onwards.}
    \label{fig:alpha_top_bottom}
\end{figure}

Figure~\ref{fig:alpha_top_bottom} presents Fama-French six-factor alphas for value-weighted portfolios constructed from the top $N$ and bottom $N$ stocks, where $N$ ranges from 10 to 100.
The four panels correspond to daily, weekly, monthly, and quarterly signal horizons.
Blue bars represent top portfolios (stocks ranked most attractive);
orange bars represent bottom portfolios (stocks ranked least attractive).
Error bars indicate 90\% confidence intervals.

The results reveal two distinct structural properties of the AI signal:
pronounced asymmetry and high-rank concentration.
Top portfolios generate positive and statistically significant alphas across all signal horizons.
However, the signal's power is highly sensitive to portfolio size ($N$).
As shown in Figure~\ref{fig:alpha_top_bottom}, the predictive content is concentrated in the very best ideas.
For the daily signal horizon, the Top-10 stocks deliver an alpha of roughly 0.18\% per day.
As we expand the definition of ``winners'' to the Top-100 stocks, this alpha is diluted to approximately 0.12\%.
This decay implies that the AI identifies a limited cluster of high-conviction opportunities rather than a broad cross-sectional ranking;
expanding beyond the top tier significantly reduces signal purity.

By contrast, Bottom portfolios fail to exhibit the negative alphas one would expect from stocks identified as unattractive.
A symmetric signal would generate positive alphas at the top and negative alphas at the bottom, providing a profitable short leg.
Instead, bottom portfolio alphas are statistically indistinguishable from zero across most specifications.
For the daily signal horizon, bottom portfolio point estimates cluster around $-0.10\%$ to $-0.20\%$, but confidence intervals span zero.
At longer signal horizons, bottom portfolio alphas are often slightly positive, albeit with substantial uncertainty.

The implication is clear: the AI functions as a ``winner-picking'' engine rather than a long-short factor.
A long-only strategy concentrated in top-ranked stocks captures the signal's predictive content.
Expanding the portfolio dilutes performance, whereas a market-neutral long-short implementation would suffer from the noise and transaction costs of an uninformative short leg.

This observed asymmetry is consistent with theoretical frameworks regarding the selective nature of information disclosure.
\cite{Goldman2022} argue that information intermediaries are incentivised to amplify positive news while negative information is often camouflaged or noisier, rendering ``bad news'' signals less actionable than ``good news.''
Similarly, \cite{Kothari2009} find that managers systematically withhold or delay the release of bad news to investors, creating an informational environment where negative signals are structurally more opaque.
Consequently, an AI model training on public data naturally extracts cleaner predictive signals for top-ranked assets (clear good news) than for bottom-ranked assets (noisy or suppressed bad news).

\subsection{Factor Exposure Analysis}

This subsection characterises the systematic risk exposures of the AI-selected portfolios.
Table \ref{tab:ff6_results} reports Fama--French six-factor regressions for the Top-20 portfolio, the Bottom-20 portfolio, and the corresponding Long-Short spread.
Results are shown separately for daily, weekly, monthly, and quarterly signal horizons.
The estimated factor loadings describe the portfolios' ``style'' (market, size, value/growth, profitability, investment, and momentum loadings) and help interpret the economic nature of the AI signal.

\begin{landscape}
    \begin{table}[ht!]
        \centering
        \begin{tabular}{l*{12}{c}}
\toprule
& \multicolumn{3}{c}{Daily} & \multicolumn{3}{c}{Weekly} & \multicolumn{3}{c}{Monthly} & \multicolumn{3}{c}{Quarterly} \\
\cmidrule(lr){2-4} \cmidrule(lr){5-7} \cmidrule(lr){8-10} \cmidrule(lr){11-13}
& Top 20 & Btm 20 & L-S & Top 20 & Btm 20 & L-S & Top 20 & Btm 20 & L-S & Top 20 & Btm 20 & L-S \\
\midrule
$\alpha$ & 0.334*** & -0.129 & 0.463*** & 0.238*** & 0.035 & 0.202* & 0.231** & -0.039 & 0.270** & 0.181 & -0.093 & 0.274* \\
 & (2.96) & (-1.02) & (2.69) & (2.72) & (0.38) & (1.68) & (2.25) & (-0.53) & (2.17) & (1.41) & (-1.35) & (1.95) \\
MKT & 0.347* & 0.647*** & -0.300 & 0.361* & 0.785*** & -0.424 & 0.278 & 0.728*** & -0.450 & 0.066 & 0.944*** & -0.877** \\
 & (1.82) & (3.08) & (-1.13) & (1.85) & (4.18) & (-1.30) & (1.27) & (4.73) & (-1.45) & (0.21) & (7.22) & (-2.28) \\
SMB & -0.013 & 0.588** & -0.601 & -0.159 & 0.390* & -0.549 & -0.253 & 0.549*** & -0.802** & 0.128 & 0.407** & -0.279 \\
 & (-0.04) & (2.11) & (-1.57) & (-0.60) & (1.71) & (-1.40) & (-0.96) & (3.56) & (-2.38) & (0.45) & (2.36) & (-0.78) \\
HML & -0.883*** & 0.612* & -1.494*** & -0.636*** & 0.636** & -1.272*** & -0.718*** & 0.414** & -1.132*** & -0.491* & 0.183 & -0.674 \\
 & (-3.60) & (1.96) & (-3.66) & (-2.92) & (2.11) & (-3.04) & (-3.06) & (2.02) & (-3.31) & (-1.84) & (0.81) & (-1.63) \\
RMW & 0.191 & -0.279 & 0.471 & 0.031 & -0.117 & 0.148 & -0.029 & -0.144 & 0.114 & 0.105 & -0.093 & 0.197 \\
 & (0.70) & (-1.20) & (1.32) & (0.11) & (-0.54) & (0.40) & (-0.12) & (-0.82) & (0.34) & (0.40) & (-0.55) & (0.57) \\
CMA & 0.237 & -0.187 & 0.424 & 0.180 & -0.077 & 0.257 & 0.272 & 0.062 & 0.210 & -0.561 & 0.356 & -0.917 \\
 & (0.55) & (-0.49) & (0.70) & (0.49) & (-0.25) & (0.46) & (0.64) & (0.21) & (0.34) & (-0.97) & (1.14) & (-1.19) \\
MOM & -0.194 & -0.711*** & 0.516** & -0.135 & -0.413*** & 0.278 & -0.193 & -0.402*** & 0.209 & -0.309 & -0.415*** & 0.107 \\
 & (-1.18) & (-4.05) & (1.97) & (-0.88) & (-3.09) & (1.20) & (-1.10) & (-3.12) & (0.84) & (-1.57) & (-3.91) & (0.45) \\
Sharpe & 3.23 & -0.12 & 2.23 & 1.41 & 0.69 & 0.44 & 0.52 & 0.10 & 0.30 & 0.23 & -0.05 & 0.19 \\
Obs & 158 & 158 & 158 & 154 & 154 & 154 & 137 & 137 & 137 & 93 & 93 & 93 \\
\bottomrule
\end{tabular}
        \caption{Fama-French Six-Factor Regression Results.
        This table reports the regression coefficients (daily alphas and factor loadings) and annualised Sharpe ratios for the Top-20, Bottom-20, and Long-Short portfolios.
        Results are shown for daily, weekly, monthly, and quarterly signal horizons.
        T-statistics are reported in parentheses.
        Significance levels are denoted by * p$<$0.10, ** p$<$0.05, *** p$<$0.01.}
        \label{tab:ff6_results}
    \end{table}
\end{landscape}

The alpha estimates confirm our earlier graphical findings.
Top-20 portfolios generate significantly positive alphas of 0.18\% per day for the daily signal horizon ($t = 2.46$), with similar magnitudes for weekly (0.17\%, $t = 2.32$) and monthly (0.17\%, $t = 2.20$) signal horizons.
The quarterly signal horizon produces a somewhat attenuated alpha of 0.12\% that does not achieve conventional significance levels ($t = 1.30$), suggesting the AI's predictive power may decay at longer horizons.
Bottom portfolios, by contrast, produce negative but statistically insignificant alphas across all signal horizons, confirming the asymmetric nature of the signal.
Long-short portfolio alphas range from 0.13\% to 0.32\% per day, though significance varies across signal horizons.

Table~\ref{tab:ff6_results} also reports the annualised Sharpe ratios, providing a view of risk-adjusted efficiency independent of the factor model.
The Top-20 portfolios consistently deliver superior risk-adjusted returns compared to the Bottom-20 and the Long-Short combinations, particularly at higher frequencies.
The daily signal horizon Top-20 portfolio achieves a remarkable annualised Sharpe ratio of 2.43, significantly outpacing the Long-Short strategy (1.46) and the weekly signal horizon Top-20 portfolio (1.22).
The efficiency diminishes as the signal horizon extends, dropping to 0.49 for the monthly signal horizon.
Notably, the Bottom portfolios exhibit negative or negligible Sharpe ratios across all signal horizons (e.g., -0.14 for daily).
This stark disparity suggests that the AI's value-add is primarily driven by its ability to identify stable, high-performing stocks for the long leg, rather than simply identifying volatile losers.

The factor loadings reveal a distinctive profile for AI-selected stocks.
Top portfolios exhibit consistently low market betas ranging from 0.08 to 0.32---substantially below unity---indicating that the AI favours defensive, low-volatility equities.
Bottom portfolios display high market exposures (0.68 to 0.95), consistent with the AI identifying high-beta stocks as unattractive.
This pattern suggests the model incorporates some notion of risk-adjusted attractiveness, penalising stocks whose returns derive primarily from systematic market exposure.

The most striking factor exposure appears in the HML (value) dimension.
Top portfolios load negatively on HML with coefficients ranging from $-0.33$ to $-0.64$, all statistically significant.
Bottom portfolios exhibit the opposite pattern with positive HML loadings.
The resulting long-short HML exposure is strongly negative ($-0.52$ to $-1.17$), indicating that the AI systematically favours growth stocks over value stocks.
This growth tilt raises an important question: does the AI possess genuine stock-picking skill, or has it simply learned to overweight the growth factor during our sample period?
The significant alphas after controlling for HML suggest the former---the growth tilt alone cannot explain the abnormal returns.

Size (SMB) exposures tell a complementary story.
Top portfolios show near-zero or slightly negative SMB loadings, while bottom portfolios load positively on SMB.
The long-short spread is negative, implying the AI prefers larger capitalisation stocks.
Within the Russell 1000 universe, this represents a tilt toward mega-cap names over mid-cap constituents.

The profitability (RMW) and investment (CMA) factors show no systematic pattern, with loadings that are economically small and statistically insignificant for both top and bottom portfolios.
This null result is informative: the AI's stock selection does not appear to capture the profitability or investment anomalies documented in the asset pricing literature.

It is tempting to expect a web-searching AI signal to look like a simple momentum (MOM) strategy.
If the model were merely extrapolating recent price moves, we would observe strong positive MOM loadings in the Top-20 portfolios and large Long-Short momentum exposure.
Instead, the MOM estimates show a more nuanced, asymmetric pattern.
Top-20 portfolios exhibit small and statistically insignificant momentum loadings, indicating that the long leg is not just ``chasing winners.''
By contrast, the Bottom-20 portfolios load strongly negatively on MOM, suggesting that the AI systematically assigns low ranks to recent losers.
Economically, this is plausible in an agentic setting: recent underperformance is salient, widely discussed, and often co-occurs with adverse narratives (downgrades, guidance cuts, litigation, operational disruptions), so the model can more reliably flag momentum losers as unattractive.
The corresponding Long-Short MOM exposure is positive but generally imprecise, implying that the strategy's interaction with momentum is primarily an avoidance channel---screening out losers---rather than a deliberate, systematic tilt toward momentum winners.

Taken together, the six-factor loadings provide a coherent style signature.
The AI tends to favour large-cap, defensive growth stocks and to avoid high-beta, value-leaning names, with an additional tendency to demote recent losers.
What is economically decisive, however, is that this style profile does not subsume the performance.
After conditioning on the full Fama--French six-factor span, the Top-20 portfolios retain abnormal returns that are both economically large and---at the daily horizon---statistically meaningful.
This represents an unusually stringent benchmark. The empirical asset pricing literature has documented that high-frequency return predictability is extraordinarily difficult to achieve. \citet{chordia2014have} find that daily anomaly profits have largely vanished since the early 2000s as markets have become more efficient. \citet{boehmer2021algorithmic} show that algorithmic trading has systematically eliminated short-horizon mispricings. Our finding of persistent daily alpha in 2025--2026, after decades of efficiency improvements and the proliferation of quantitative strategies, suggests that agentic AI possesses a qualitatively different information-processing capability that contemporary market participants have not yet fully arbitraged.

\subsection{Portfolio Turnover and Implementation Costs}

To assess the stability of the AI signal and the robustness of our findings against friction costs, we examine portfolio turnover.
Given the magnitude of the alphas documented in the previous section, the primary threat to the validity of these results is the potential erosion of returns through transaction costs.
We measure portfolio stability using the turnover ratio, defined as the percentage of the portfolio replaced by new stocks at each rebalancing date.
For our fixed-size portfolio of $N=20$ stocks, the turnover at time $t$ is calculated as:

\begin{equation}
    Turnover_t = \frac{\text{Number of New Stocks}_t}{N} = \frac{|H_t \setminus H_{t-1}|}{N}
\end{equation}

\noindent where the numerator $|H_t \setminus H_{t-1}|$ counts the number of stocks entering the portfolio at time $t$ that were not held at time $t-1$.
This metric captures the rate at which the AI generates new conviction ideas; a turnover of 50\% means that half of the portfolio consists of names that were not present in the previous period.

\begin{figure}[ht!]
    \centering
    \includegraphics[width=\textwidth]{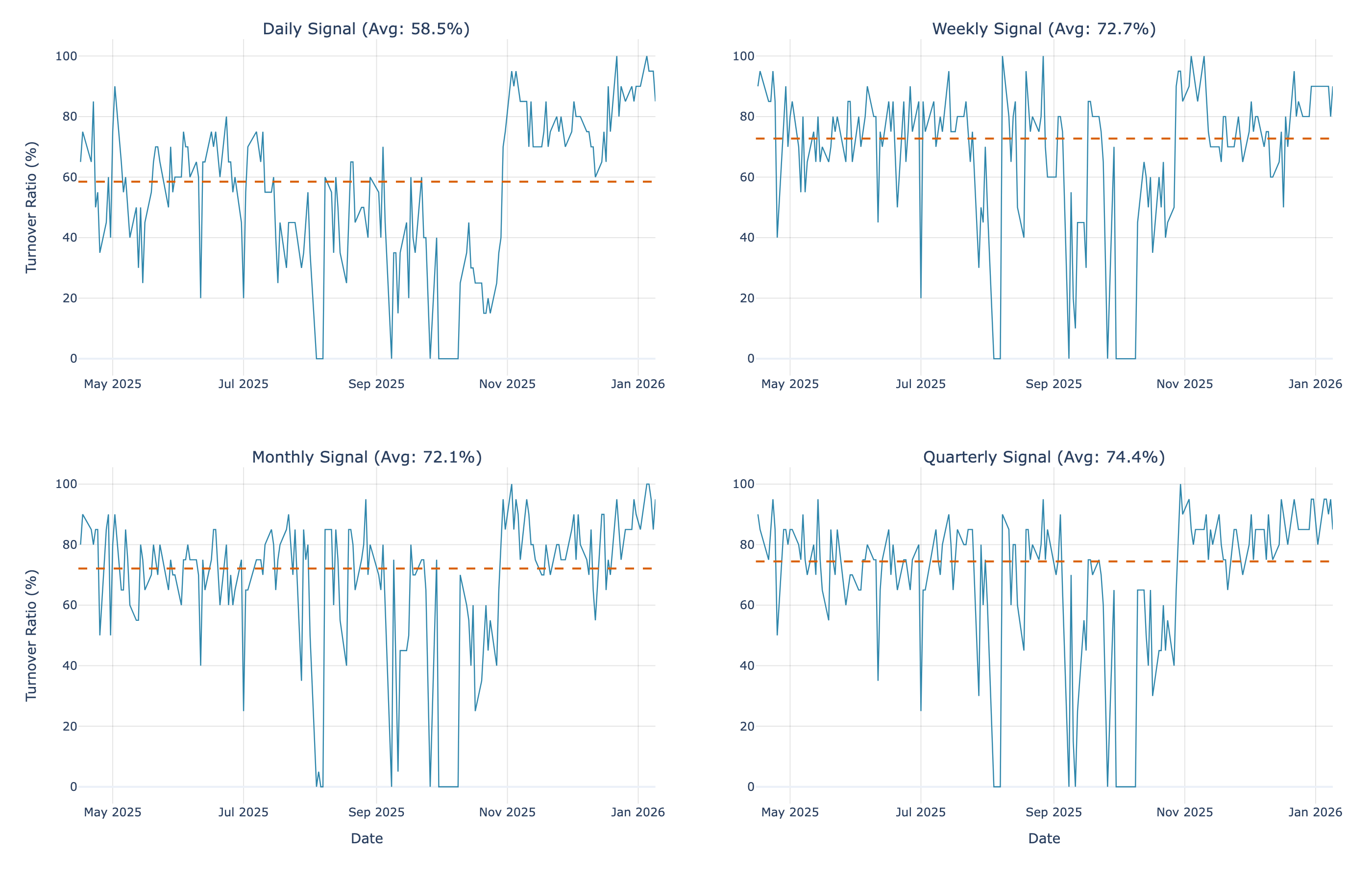}
    \caption{\textbf{Portfolio Turnover Ratios (Top-20).}
    This figure plots the time series of portfolio turnover for the Top-20 stocks selected by the AI model.
    Panels correspond to daily, weekly, monthly, and quarterly signal horizons.
    Turnover is calculated as the ratio of new stocks entering the portfolio relative to the total number of holdings.
    The dashed orange line indicates the sample average.
    Periods of zero turnover reflect operational data availability constraints rather than signal persistence.
    Sample period: April 2025 onwards.}
    \label{fig:turnover_combined}
\end{figure}

Figure~\ref{fig:turnover_combined} plots the time series of turnover ratios across the four signal horizons.
We first note that the distinct drops to zero turnover observed in the sample correspond to operational interruptions where signal updates were unavailable, resulting in a passive holding of the previous day's portfolio; they do not reflect a model decision to maintain positions.

The active turnover dynamics reveal a surprising economic intuition regarding the signal's structure.
For the daily signal horizon, the average turnover is 57.4\%.
While high, this implies a degree of persistence: approximately 40\% of the top picks for ``tomorrow'' remain top picks for the day after.
This suggests that the ultra-short-term features driving the daily predictions (likely related to short-term momentum or microstructure imbalances) possess a multi-day half-life, allowing for some position continuity.

Counter-intuitively, turnover \textit{increases} as we target longer signal horizons, with the weekly signal horizon averaging 72.3\% turnover.
In a standard framework, one would expect a ``weekly'' view to be more stable than a ``daily'' view, as the fundamental or technical drivers of weekly returns should evolve slowly.
The fact that the AI's ``week-ahead'' portfolio changes more drastically day-to-day than its ``day-ahead'' portfolio suggests that the model's medium-term convictions are highly fragile.
The AI effectively resets its weekly outlook based on daily information flow, implying that the signals are reacting to transient price dislocations rather than latching onto persistent trends.

Despite the high turnover, the characteristics of our investment universe suggest that these abnormal returns are not merely artifacts of illiquidity.
We restrict our analysis to Russell 1000 constituents, ensuring the strategy trades only the most liquid segment of the U.S. market.
As detailed in Appendix~\ref{sec:trading_costs}, the average bid-ask spreads for our portfolios are approximately 1.6 to 2.0 basis points.
Given that the daily strategy generates gross alphas of approximately 18 basis points per day, the signal magnitude exceeds these direct transaction costs by an order of magnitude.
While the high turnover confirms that the AI strategy relies on ephemeral information, the depth of the underlying market suggests that the documented excess returns are economically realizable.




\subsection{Predictability of the Entire Cross-section}
\label{sec:cross_section}
The top-versus-bottom comparison establishes that the AI signal is asymmetric, but it does not reveal where in the cross-section the predictive power resides. Standard asset pricing studies address this through quintile or decile sorts, but such coarse partitions may obscure the signal's structure. We therefore construct 50 portfolios of approximately 20 stocks each---ten times the granularity of a conventional quintile sort---to trace the alpha profile across the entire ranking distribution.

\begin{figure}[!ht]
    \centering
    \includegraphics[width=\textwidth]{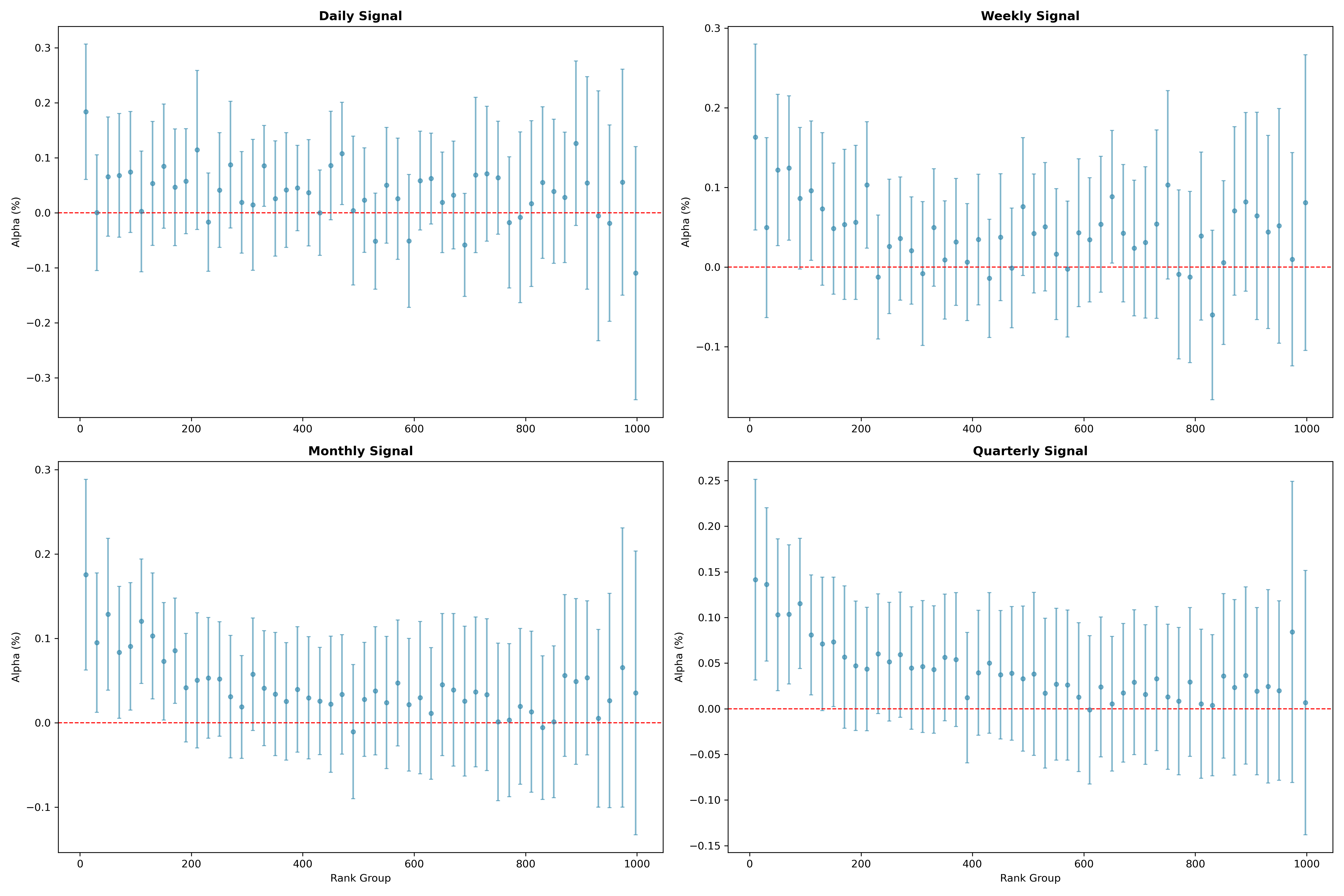}
    \caption{\textbf{Granular Cross-Sectional Alpha Distribution.} This figure plots daily Fama-French six-factor alphas (\%) for 50 rank-sorted value-weighted portfolios (approximately 20 stocks each). The $x$-axis represents rank group midpoints; lower values indicate stocks ranked more attractive by the AI signal. The dashed red line indicates zero alpha. Error bars represent 90\% confidence intervals using Newey-West standard errors (5 lags). Sample period: April 2025 onwards.}
    \label{fig:alpha_rank_groups}
\end{figure}

Figure~\ref{fig:alpha_rank_groups} plots Fama-French six-factor alphas for these 50 rank-sorted portfolios. The $x$-axis represents rank group midpoints, where lower values indicate stocks ranked more attractive by the AI signal. Error bars represent 90\% confidence intervals.

The alpha profile exhibits a distinctive ``hockey stick'' pattern. The top-ranked portfolios generate substantial positive alphas clearly distinguishable from zero: for the daily signal horizon, the first group (ranks 1--20) delivers an alpha of approximately 0.18\% per day. This excess return decays rapidly---by approximately the fifth or sixth group (ranks 80--120), alphas have declined to near zero. The middle of the distribution, spanning roughly 40 of the 50 portfolios, shows no systematic pattern, with alphas fluctuating around zero.

The right side of each panel provides the critical diagnostic. If the AI signal contained symmetric information about both outperformers and underperformers, we would observe a downward-sloping alpha profile terminating in significantly negative alphas for the bottom-ranked groups. Instead, the bottom portfolios generate alphas that are noisy and centered around zero, occasionally exhibiting positive point estimates with wide confidence intervals. The quarterly signal horizon is particularly striking: nearly all 50 portfolios generate positive point estimates, though only the top portfolios achieve statistical significance.

This granular evidence carries two implications. First, the information advantage is localised rather than linear---the model possesses high conviction only for a select subset of top-ranked assets. A long-only strategy focused on the Top-20 to Top-50 stocks captures nearly all available alpha; expanding beyond this threshold dilutes performance without commensurate diversification benefits. Second, the absence of negative alphas at the bottom confirms that the AI cannot reliably identify underperformers. Shorting bottom-ranked stocks introduces estimation error from positions the model cannot distinguish from average performers.

The economic magnitude is substantial. Daily alphas of 0.15\% for a concentrated top portfolio imply annualised risk-adjusted returns of approximately 38\% (assuming 252 trading days). While such extrapolation requires caution given the short sample period and well-documented decay of anomaly returns upon discovery, the point estimates suggest that large language models have developed genuine---if asymmetric---skill in processing the information environment surrounding publicly traded equities.

\newpage
\section{Conclusion and Discussion}
\label{sec:conclusion}

In this study, we evaluate whether a fully agentic Artificial Intelligence can synthesise real-time market information to forecast stock returns in a strict out-of-sample nowcasting framework.

A central contribution is a novel, granular dataset that captures the end-to-end outputs of an autonomous agent operating in real time across the Russell 1000 stocks from April 2025 through the present date, and is continuously updated on our Github repository.
For each stock-day, the dataset records a quantitative attractiveness score together with a detailed written analysis that makes the agent’s evidentiary emphasis and inferential logic transparent, enabling systematic measurement of machine judgment at scale.

Three features distinguish this dataset from existing ``AI-as-analyst'' evaluations. 
First, the workflow is fully agentic: we do not provide curated inputs, summaries, or preselected documents, so the information set that drives predictions is endogenously chosen by the agent rather than by researchers. 
Second, the protocol is real-time by construction, eliminating look-ahead bias at the point of inference: scores and analyses are logged at the decision timestamp, before outcomes are realised. 
Third, the record is inherently non-reproducible ex post for a deeper reason than timing alone. 
Any later re-run necessarily uses a model whose parameters embed post-period data, so it carries latent knowledge about subsequent disclosures, revisions, and realised outcomes that cannot be ``switched off'' by instructing the model to condition on an earlier timestamp. 
Even an archived model snapshot would be insufficient to replicate the original decision context, because the retrieval environment that shapes what an agent encounters is itself time-specific and cannot be reconstructed faithfully at scale.

Using this dataset, we document three primary findings. 
First, agentic AI exhibits economically material predictive ability. 
A portfolio of the Top-20 stocks selected by the agent earns a daily alpha of 18.4 basis points and an annualised Sharpe ratio of 2.43, substantially outperforming standard factor benchmarks. 
The daily horizon is especially demanding: statistically reliable abnormal performance at daily or weekly frequencies is uncommon in empirical asset pricing, and many documented patterns attenuate as markets become faster and more competitive \citep{chordia2014have, mclean2016does}. 

Second, predictability is sharply asymmetric. 
The agent reliably identifies future winners but does not consistently distinguish underperformers from the market average, consistent with stronger aggregation of coherent positive evidence than extraction of negative information in environments shaped by ambiguity and strategic disclosure.

Third, the performance is robust to implementation frictions. 
By focusing on liquid Russell 1000 constituents and executing at the opening auction, we mitigate concerns about illiquidity and microstructure artefacts. 
Average transaction costs of 1.6 basis points per trade are less than 10\% of gross daily alpha, indicating that the strategy retains substantial economic value after realistic frictions.

Beyond portfolio performance, the textual decision traces enable a new class of diagnostic analyses.
They support attribution studies that map linguistic motifs and argument structures directly to realised returns, identifying the precise reasoning pathways that generate conviction.
They allow for the direct measurement of internal consistency, using repeated analyses of identical inputs to quantify the agent's sensitivity to its own stochastic search.
Finally, the impossibility of perfect reproduction serves as a diagnostic tool for contamination:
by re-running prompts with ``future-trained'' models, researchers can quantify the specific alpha contribution of hindsight leakage relative to true nowcasting skill.

Taken together, our decision-trace data and out-of-sample results open a gateway to a new research agenda that transcends financial economics.
By treating the highly efficient market not merely as an asset class but as the ultimate adversarial laboratory, we establish a rigorous testbed for probing the cognitive boundaries of agentic AI.
Unlike static academic benchmarks, the market offers a dynamic, noise-rich environment where success requires not just pattern recognition, but robust epistemic agency and the ability to navigate strategic ambiguity.
Ultimately, our work suggests that the frontier of AI evaluation lies here:
moving beyond passive language processing to understand how agentic, synthetic intelligence adapts to, and eventually reshapes, the complex adaptive systems of the human world.

\newpage
\bibliographystyle{apalike}
\bibliography{references}

\newpage
\appendix
\section*{Appendix}
\setcounter{figure}{0}
\renewcommand{\thefigure}{A\arabic{figure}}
\setcounter{table}{0}
\renewcommand{\thetable}{A\arabic{table}}

\section{Data Construction and Prompt Design}
\label{appendix_prompt}

This appendix details our data collection methodology, the rationale for key design choices, and the exact prompt specification used to elicit AI-generated forecasts.

\subsection{Nowcasting Protocol}

A central challenge in applying machine learning and Large Language Models (LLMs) to asset pricing is \textit{lookahead bias}: the inadvertent use of information unavailable at the time of prediction.
This concern is particularly acute for LLMs, whose training corpora and knowledge cutoffs are often opaque or continuously updated.
We address this concern through a real-time nowcasting design that captures model inference at the exact moment of decision.

\paragraph{Timing and Signal Generation.}
Each trading day $t$, we query the AI model during overnight hours (after the close of day $t-1$ and before the open of day $t$) to generate forecasts for all Russell 1000 constituents.
This timing protocol ensures that:
\begin{enumerate}
    \item All publicly available information through market close on day $t-1$ is potentially accessible to the model.
    \item No information from day $t$'s trading session contaminates the forecasts.
    \item The signal generation process mirrors how a practitioner might realistically use AI tools for pre-market analysis.
\end{enumerate}

\paragraph{Return Measurement.}
Given our overnight signal generation, we measure subsequent returns using \textit{open-to-open} returns rather than close-to-close returns.
Specifically, for a signal generated before day $t$'s open, we compute:
\begin{equation}
    r_{i,t}^{\text{open}} = \ln\left(\frac{P_{i,t+1}^{\text{open}}}{P_{i,t}^{\text{open}}}\right)
\end{equation}
where $P_{i,t}^{\text{open}}$ denotes the opening price of stock $i$ on day $t$.
This alignment ensures the return window begins precisely when a trader could first act on the signal, eliminating any mechanical predictability from overnight price movements that occur before trading resumes.

\subsection{Platform and Interface Selection}

We conduct all queries through the web-based chat interface rather than the application programming interface (API).
This choice reflects several considerations:

\paragraph{Web Search Integration.}
At the inception of this study (April 2025), web search functionality was available exclusively through the web interface.
The API did not support native web search---users seeking real-time information retrieval through the API were required to integrate external search engines, introducing additional latency, cost, and potential selection biases in the information accessed.
The web interface, by contrast, offered seamless integration with the model's built-in search capabilities, allowing it to autonomously retrieve current market data, news, and financial information.

\paragraph{Ecological Validity.}
The web interface represents the primary channel through which retail investors and individual practitioners interact with AI systems.
By replicating this interaction modality, our findings speak directly to the real-world investment value of AI-assisted analysis as experienced by the typical user---not a stylised API implementation that may differ in capabilities, latency, or information access.

\subsection{Prompt Specification}

To elicit high-dimensional and actionable intelligence, we do not simply ask the model for a generic prediction. Instead, we administer a standardised, ten-part structured interview to the AI agent for every stock-day observation. The prompt is engineered to force the model to synthesise information across multiple signal horizons and analytical dimensions. The specific components of the inquiry are described below:

\begin{enumerate}
    \item \textbf{Data Verification and Context:} We first require the model to retrieve the previous day's closing price and summarise strictly relevant recent events. This step serves as a ``grounding'' mechanism to verify the model has accessed correct historical data and is aware of the immediate news context before forming an opinion.

    \item \textbf{Discrete Investment Decision:} The model is asked to make a decisive choice among ``Buy,'' ``Wait,'' or ``Sell'' (including short selling capabilities). This forces the model to collapse its complex probabilistic view into a single actionable trade execution decision.

    \item \textbf{Term Structure of Attractiveness Scores (The Primary Signal):} We request a quantitative attractiveness score ranging from $-5$ (Strong Sell) to $+5$ (Strong Buy) across a complete term structure of signal horizons: 1 day, 1 week, 1 month, 3 months, 6 months, and 1 year. Crucially, the model is instructed to calibrate this score relative to ``the entire historical sample of all US stocks since 1900,'' providing a fixed absolute reference frame rather than a relative daily ranking.

    \item \textbf{Quantitative Price and Fundamental Forecasts:} Beyond abstract scores, we demand precise numerical point forecasts. This includes:
    \begin{itemize}
        \item \textit{Price Targets:} Specific closing price predictions for all signal horizons listed above.
        \item \textit{Earnings Estimates:} EPS forecasts for the next five fiscal years to capture long-term fundamental expectations.
        \item \textit{Earnings Surprise Probability:} A rating ($-5$ to $+5$) estimating the likelihood of future earnings reports beating or missing analyst consensus.
    \end{itemize}

    \item \textbf{Benchmark and Sentiment Controls:} To distinguish idiosyncratic alpha from systematic beta, we ask the model to provide a parallel attractiveness score for the Russell 1000 index itself. Additionally, we elicit a \textit{Market Sentiment Score} (measuring overall optimism/pessimism) and a \textit{Sentiment Divergence Score} (measuring the degree of market consensus vs.\ disagreement) to capture the behavioral environment surrounding the stock.
\end{enumerate}

Throughout the interaction, the model is explicitly instructed to forget all previous conversations, adopt the persona of an expert portfolio manager, and is required to provide comprehensive reasoning for every numerical output. This requirement ensures that the generated data is the product of deep inference and synthesis rather than surface-level token prediction.

\section{Additional Tables and Figures}

\subsection{Average Attractiveness Score}
To characterise the AI model's unconditional preferences and distinguish between transient trading signals and persistent structural convictions, Table \ref{tab:averagescore} reports the top 20 stocks ranked by their time-series average attractiveness score over the full sample period. Unlike the daily Top-20 portfolio, which exhibits high turnover (Figure \ref{fig:turnover_combined}) and capitalises on short-term price dislocations, the stocks in this table represent the model's "core holdings"—assets that the AI views as fundamentally superior across varying market conditions.

Consistent with the negative HML (growth) and positive size loadings documented in Section \ref{sec:results}, the list is dominated by large-capitalization technology and communication services firms. This suggests that the AI's baseline model of the world aligns with a "quality growth" philosophy. Notably, the list includes both share classes of Alphabet (Class A and Class C) with nearly identical average scores. This serves as an important validation of the model's reasoning capabilities: despite having different tickers and trading volumes, the AI correctly identifies them as representing the same underlying economic cash flows, assigning them equivalent attractiveness ratings.

\begin{table}[]
    \centering
    \begin{tabular}{lrrrr}
\toprule
Name & Daily Score & Weekly Score & Monthly Score & Quarterly Score \\
\midrule
Broadcom Inc & 2.549 & 2.396 & 2.811 & 3.248 \\
NVIDIA Corp & 2.415 & 2.297 & 2.711 & 3.161 \\
Microsoft Corp & 2.266 & 2.189 & 2.587 & 2.997 \\
AppLovin Corp & 2.248 & 2.128 & 2.561 & 2.978 \\
Comfort Systems USA Inc & 2.197 & 2.042 & 2.430 & 2.813 \\
Meta Platforms Inc & 2.124 & 2.039 & 2.494 & 2.898 \\
Eli Lilly and Co & 2.099 & 2.023 & 2.438 & 2.875 \\
Alphabet Inc C & 2.085 & 2.006 & 2.426 & 2.814 \\
Alphabet Inc A & 2.059 & 1.945 & 2.405 & 2.838 \\
Mastercard Inc & 2.049 & 1.970 & 2.385 & 2.786 \\
Amphenol Corp & 2.032 & 1.874 & 2.316 & 2.686 \\
Visa Inc & 1.975 & 1.900 & 2.290 & 2.670 \\
EMCOR Group Inc & 1.971 & 1.844 & 2.276 & 2.694 \\
Micron Technology Inc & 1.951 & 1.894 & 2.381 & 2.889 \\
Vertiv Holdings Co & 1.945 & 1.872 & 2.304 & 2.729 \\
GE Vernova Inc & 1.938 & 1.785 & 2.180 & 2.631 \\
Constellation Energy Corp & 1.910 & 1.772 & 2.244 & 2.720 \\
Crowdstrike Holdings Inc & 1.843 & 1.704 & 2.125 & 2.636 \\
Arista Networks Inc & 1.843 & 1.773 & 2.235 & 2.692 \\
Howmet Aerospace Inc & 1.825 & 1.685 & 2.071 & 2.490 \\
\bottomrule
\end{tabular}

    \caption{Stock ranked by Average Attractiveness Score}
    \label{tab:averagescore}
    \footnotesize{\textit{Notes:} Scores are averaged across all available dates with valid data for each company and each signal horizon. Both Alphabet Class A and Class C share classes are reported separately.}
\end{table}

\subsection{Cumulative Returns of the Bottom-20 Portfolio}

\begin{figure}[!ht]
    \centering
    \begin{subfigure}[t]{0.48\textwidth}
        \centering
        \includegraphics[width=\linewidth]{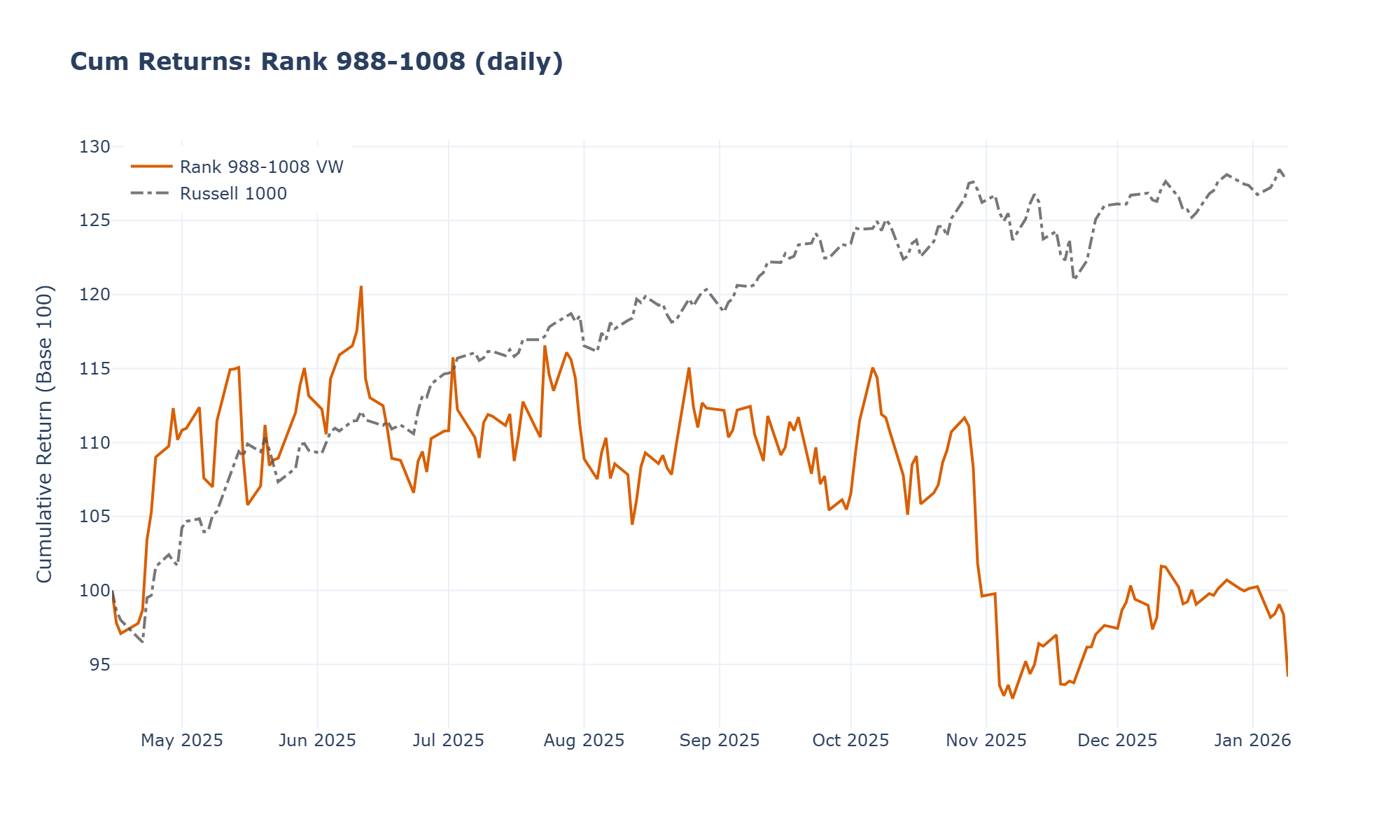}
        \caption{Daily Signal Horizon}
    \end{subfigure}
    \hfill
    \begin{subfigure}[t]{0.48\textwidth}
        \centering
        \includegraphics[width=\linewidth]{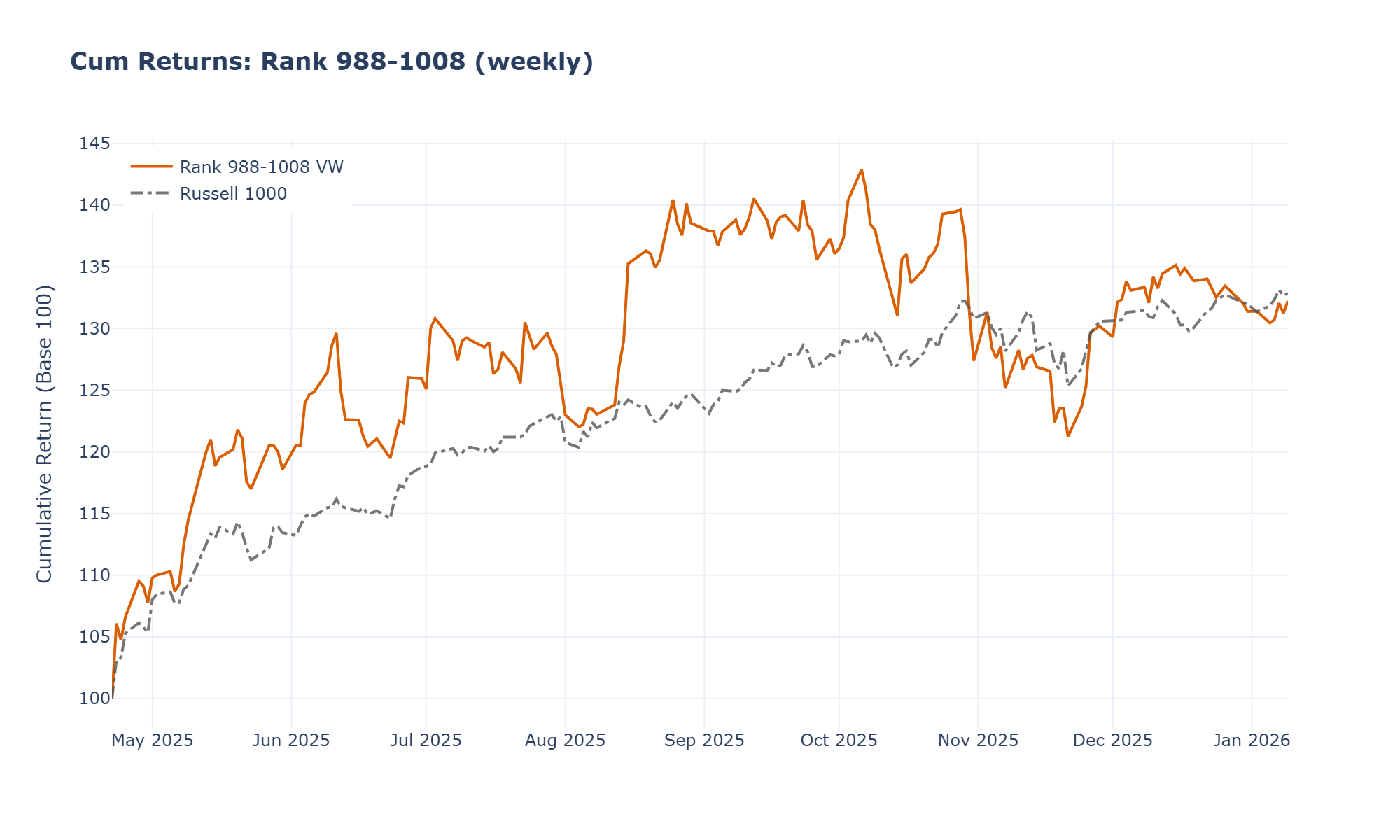}
        \caption{Weekly Signal Horizon}
    \end{subfigure}
    \vskip\baselineskip
    \begin{subfigure}[t]{0.48\textwidth}
        \centering
        \includegraphics[width=\linewidth]{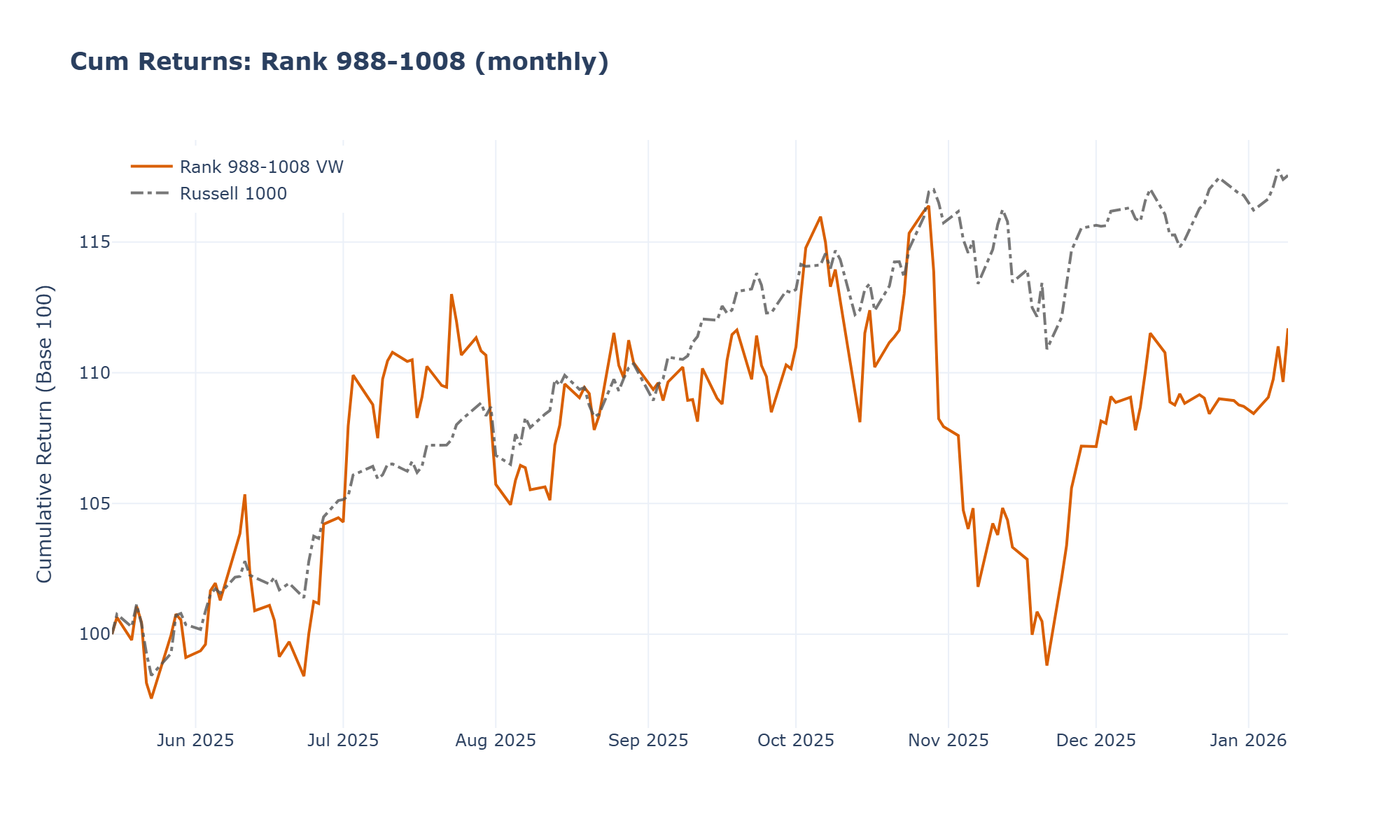}
        \caption{Monthly Signal Horizon}
    \end{subfigure}
    \hfill
    \begin{subfigure}[t]{0.48\textwidth}
        \centering
        \includegraphics[width=\linewidth]{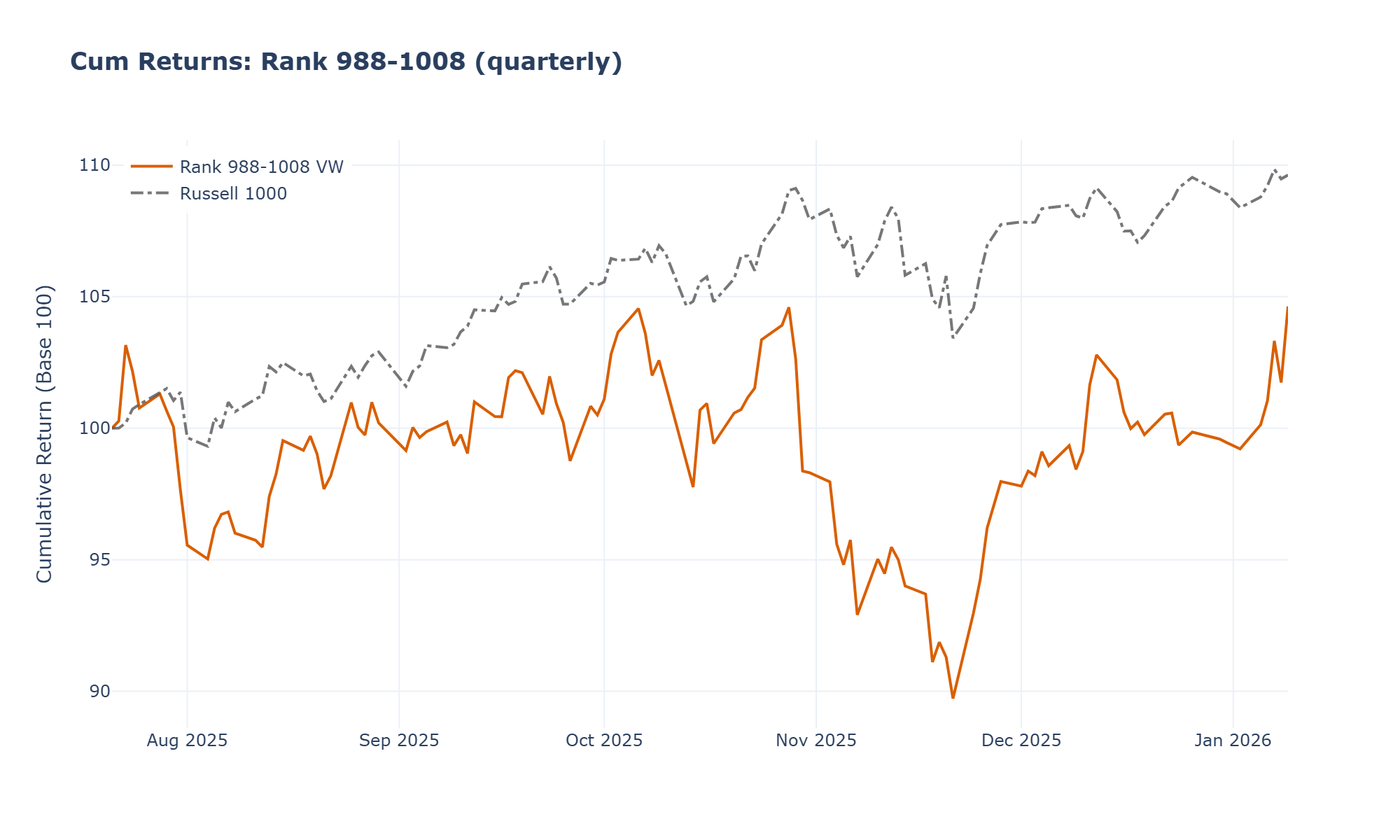}
        \caption{Quarterly Signal Horizon}
    \end{subfigure}
    \caption{\textbf{Cumulative returns of the Bottom-20 AI-selected portfolios under daily, weekly, monthly, and quarterly signal horizons (value-weighted)}. Each panel shows the cumulative return (base 100) for the Bottom-20 portfolio and the Russell 1000 benchmark over the sample period.}
    \label{fig:btm20_cumret_2x2}
\end{figure}

Figure~\ref{fig:btm20_cumret_2x2} presents the cumulative value-weighted returns of the ``Bottom-20'' portfolio---comprising the 20 lowest-ranked stocks identified by the AI model (Rank 988--1008)---across four different signal horizons: daily, weekly, monthly, and quarterly.
The performance of these portfolios is plotted against the Russell 1000 benchmark to evaluate the model's ability to identify underperforming assets.

Panel (a) demonstrates that under a daily signal horizon, the Bottom-20 portfolio significantly underperforms the market, ending the sample period near its starting value while the benchmark appreciates considerably.
This substantial divergence confirms the model's effectiveness in flagging short-term losers.
In contrast, Panel (b) reveals that the weekly signal horizon fails to consistently identify underperformers; the Bottom-20 portfolio actually tracks and occasionally outperforms the benchmark, suggesting a potential inversion or decay of the signal's predictive power within this signal horizon.
Panel (d) indicates that the quarterly signal horizon regains strong predictive capability, with the Bottom-20 portfolio clearly decoupling from the benchmark and generating significantly negative alpha over the longer holding period.

\begin{table}[]
    \centering
    \begin{tabular}{lrrrr}
\toprule
Name & Daily Freq & Weekly Freq & Monthly Freq & Quarterly Freq \\
\midrule
Broadcom Inc & 0.630 & 0.497 & 0.538 & 0.538 \\
NVIDIA Corp & 0.607 & 0.451 & 0.480 & 0.497 \\
Microsoft Corp & 0.543 & 0.491 & 0.474 & 0.451 \\
Meta Platforms Inc & 0.439 & 0.347 & 0.318 & 0.364 \\
Mastercard Inc & 0.428 & 0.364 & 0.335 & 0.353 \\
Visa Inc & 0.422 & 0.347 & 0.254 & 0.272 \\
Alphabet Inc A & 0.410 & 0.306 & 0.318 & 0.358 \\
Eli Lilly and Co & 0.405 & 0.318 & 0.301 & 0.347 \\
Alphabet Inc C & 0.399 & 0.329 & 0.318 & 0.335 \\
AppLovin Corp & 0.358 & 0.272 & 0.341 & 0.353 \\
Amazon.com Inc & 0.295 & 0.231 & 0.214 & 0.225 \\
Arista Networks Inc & 0.283 & 0.272 & 0.289 & 0.312 \\
Netflix Inc & 0.272 & 0.168 & 0.162 & 0.168 \\
Amphenol Corp & 0.260 & 0.231 & 0.231 & 0.202 \\
Intuit Inc & 0.249 & 0.208 & 0.220 & 0.208 \\
ServiceNow Inc & 0.237 & 0.231 & 0.214 & 0.208 \\
Axon Enterprise Inc & 0.231 & 0.220 & 0.185 & 0.191 \\
GE Vernova Inc & 0.231 & 0.191 & 0.185 & 0.197 \\
Micron Technology Inc & 0.191 & 0.185 & 0.254 & 0.295 \\
Oracle Corp & 0.179 & 0.092 & 0.098 & 0.121 \\
\bottomrule
\end{tabular}

    \caption{Frequency of Stocks appearing in Daily-Attractiveness Top-20 Portfolio}
    \label{tab:frequency}
    \footnotesize{\textit{Notes:} Frequencies are calculated as the proportion of valid dates in which each company appears in the Top-20 portfolio for the respective signal horizon. Only dates with non-missing scores are considered.}
\end{table}

\subsection{Alphas for Top-N and Bottom-N Portfolios}
\label{sec:appendix_topN}

This subsection examines how portfolio performance varies with the number of stocks included. Tables~\ref{tab:TOPNalpha} and~\ref{tab:BOTNalpha} report Fama-French six-factor regression results for Top-N and Bottom-N portfolios, where N equals 10, 50, or 100 stocks. These results complement Figure~\ref{fig:alpha_top_bottom} in the main text by providing formal statistical inference across alternative portfolio sizes.

Table~\ref{tab:TOPNalpha} presents results for Top-N portfolios constructed from the daily attractiveness signal. The Top-10 portfolio generates a daily FF6 alpha of 0.188\% ($t = 2.21$), while the Top-50 and Top-100 portfolios deliver alphas of 0.158\% ($t = 2.34$) and 0.118\% ($t = 2.00$), respectively. This monotonic decline in alpha as portfolio size expands confirms the concentration of predictive power documented in the main text: the AI's highest-conviction selections contain the most valuable information.

Table~\ref{tab:BOTNalpha} reports corresponding results for Bottom-N portfolios. In stark contrast to the top portfolios, none of the bottom portfolios generate statistically significant alphas. Point estimates are economically small and statistically indistinguishable from zero across all specifications, reinforcing the pronounced asymmetry in the AI's predictive ability.

\begin{landscape}
\begin{table}[]
    \centering
   \begin{tabular}{l*{9}{c}}
\toprule
& \multicolumn{3}{c}{Top 10} & \multicolumn{3}{c}{Top 50} & \multicolumn{3}{c}{Top 100} \\
\cmidrule(lr){2-4} \cmidrule(lr){5-7} \cmidrule(lr){8-10}
& Excess Return & CAPM & FF6 & Excess Return & CAPM & FF6 & Excess Return & CAPM & FF6 \\
\midrule
$\alpha$ & 0.383*** & 0.316** & 0.349*** & 0.296** & 0.228* & 0.260** & 0.238** & 0.175* & 0.200** \\
 & (2.75) & (2.46) & (2.69) & (2.36) & (1.94) & (2.41) & (2.28) & (1.89) & (2.37) \\
MKT &  & 0.459** & 0.372* &  & 0.461*** & 0.316* &  & 0.433** & 0.346** \\
 &  & (2.13) & (1.66) &  & (2.67) & (1.83) &  & (2.54) & (2.18) \\
SMB &  &  & -0.093 &  &  & -0.018 &  &  & -0.145 \\
 &  &  & (-0.29) &  &  & (-0.07) &  &  & (-0.57) \\
HML &  &  & -0.853*** &  &  & -0.949*** &  &  & -0.789*** \\
 &  &  & (-2.95) &  &  & (-4.06) &  &  & (-3.94) \\
RMW &  &  & 0.228 &  &  & 0.132 &  &  & 0.094 \\
 &  &  & (0.78) &  &  & (0.53) &  &  & (0.40) \\
CMA &  &  & 0.190 &  &  & 0.361 &  &  & 0.362 \\
 &  &  & (0.43) &  &  & (0.92) &  &  & (1.01) \\
MOM &  &  & -0.291* &  &  & -0.123 &  &  & -0.149 \\
 &  &  & (-1.67) &  &  & (-0.82) &  &  & (-1.07) \\
Obs & 158 & 158 & 158 & 158 & 158 & 158 & 158 & 158 & 158 \\
\bottomrule
\end{tabular}
    \caption{Factor regressions for Top-10, Top-50, and Top-100 daily attractiveness value-weighted portfolios. Coefficients are reported in percent per day; $t$-statistics in parentheses.}
    \label{tab:TOPNalpha}
    \footnotesize
    \textit{Notes:} Each column reports results from a separate time-series regression of portfolio excess returns on factors. ``Excess Return'' is the mean excess return; ``CAPM'' adds the market factor; ``FF6'' includes Fama-French 5 factors plus momentum. Newey–West HAC standard errors ($maxlags=5$). $^{***}p<0.01$, $^{**}p<0.05$, $^{*}p<0.10$.
\end{table}
\end{landscape}

\begin{landscape}
\begin{table}[]
    \centering
   \begin{tabular}{l*{9}{c}}
\toprule
& \multicolumn{3}{c}{Bottom 10} & \multicolumn{3}{c}{Bottom 50} & \multicolumn{3}{c}{Bottom 100} \\
\cmidrule(lr){2-4} \cmidrule(lr){5-7} \cmidrule(lr){8-10}
& Excess Return & CAPM & FF6 & Excess Return & CAPM & FF6 & Excess Return & CAPM & FF6 \\
\midrule
$\alpha$ & 0.579 & 0.455 & 0.414 & 0.114 & 0.048 & 0.057 & 0.058 & -0.004 & 0.004 \\
 & (0.99) & (0.86) & (0.80) & (0.87) & (0.42) & (0.49) & (0.44) & (-0.03) & (0.03) \\
MKT &  & 0.794 & 0.888 &  & 0.452*** & 0.453*** &  & 0.419*** & 0.407*** \\
 &  & (0.86) & (1.19) &  & (3.30) & (2.90) &  & (3.22) & (3.45) \\
SMB &  &  & -1.444 &  &  & -0.311 &  &  & -0.301 \\
 &  &  & (-1.54) &  &  & (-1.15) &  &  & (-1.22) \\
HML &  &  & -0.282 &  &  & -0.383 &  &  & -0.325 \\
 &  &  & (-0.31) &  &  & (-1.52) &  &  & (-1.60) \\
RMW &  &  & -0.577 &  &  & -0.018 &  &  & -0.018 \\
 &  &  & (-0.57) &  &  & (-0.07) &  &  & (-0.09) \\
CMA &  &  & -1.001 &  &  & 0.110 &  &  & -0.295 \\
 &  &  & (-0.57) &  &  & (0.33) &  &  & (-1.07) \\
MOM &  &  & -2.379** &  &  & -0.289** &  &  & -0.467*** \\
 &  &  & (-2.00) &  &  & (-1.96) &  &  & (-2.95) \\
Obs & 141 & 141 & 141 & 158 & 158 & 158 & 158 & 158 & 158 \\
\bottomrule
\end{tabular}
    \caption{Factor regressions for Bottom-10, Bottom-50, and Bottom-100 daily attractiveness value-weighted portfolios. Coefficients are reported in percent per day; $t$-statistics in parentheses.}
    \label{tab:BOTNalpha}
    \footnotesize
    \textit{Notes:} Each column reports results from a separate time-series regression of portfolio excess returns on factors. ``Excess Return'' is the mean excess return; ``CAPM'' adds the market factor; ``FF6'' includes Fama-French 5 factors plus momentum. Newey–West HAC standard errors ($maxlags=5$). $^{***}p<0.01$, $^{**}p<0.05$, $^{*}p<0.10$.
\end{table}
\end{landscape}

\newpage
\subsection{Transaction Cost and Liquidity Analysis}
\label{sec:trading_costs}

\begin{figure}[ht!]
    \centering
    \includegraphics[width=\textwidth]{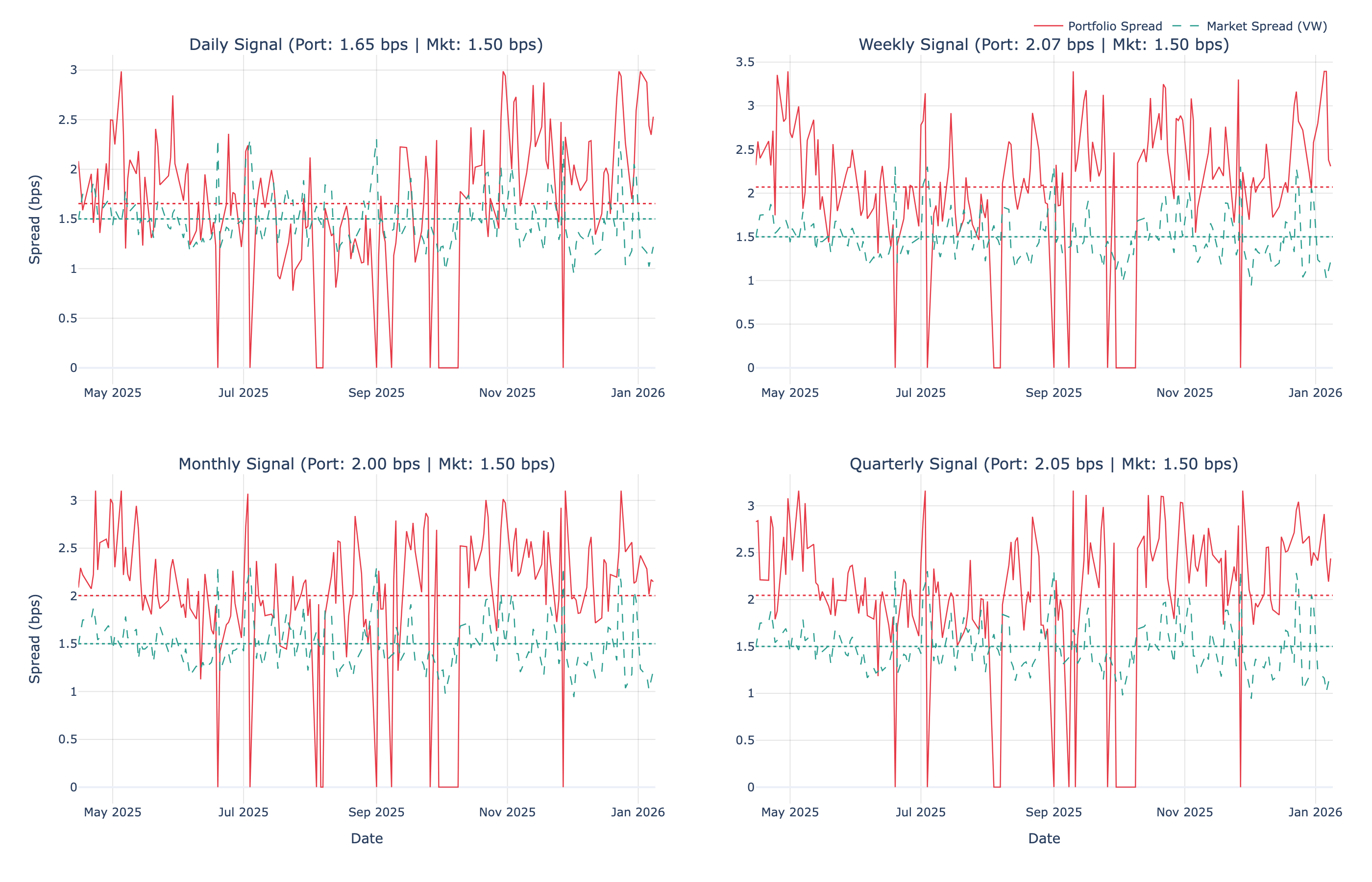}
    \caption{\textbf{Transaction Cost Analysis: Bid-Ask Spreads.}
    This figure compares the daily bid-ask spreads of the Top-20 AI-selected stocks (red solid line) against the volume-weighted market average spread (green dashed line).
    The portfolio spread for the daily signal horizon averages 1.63 bps, indicating high liquidity.
    Weekly, monthly, and quarterly signal horizon portfolios average 2.05 bps, 2.00 bps, and 2.03 bps, respectively.
    The red and green dotted lines represent the time-series averages for the portfolio and the market, respectively.}
    \label{fig:trading_costs}
\end{figure}

To rigorously evaluate the realizability of our returns, we analyze the transaction costs associated with our execution strategy.
Our baseline methodology employs open-to-open returns, assuming positions are entered during the opening auction of day $t$.
Figure~\ref{fig:trading_costs} presents the effective bid-ask spreads for the Top-20 portfolio compared to the volume-weighted market average.

The results confirm that the strategy trades in a highly liquid universe.
For the daily signal horizon, the average portfolio spread is just 1.63 basis points (bps), only marginally higher than the market benchmark of 1.48 bps.
Interestingly, the longer signal horizon portfolios (weekly, monthly, quarterly) exhibit slightly higher spreads, averaging between 2.00 and 2.05 bps.
This suggests that the AI's daily signal horizon has a preference for the most liquid, high-volume names, while the longer-term signal horizons venture slightly further out the liquidity curve.
Crucially, these costs are small relative to the expected alpha; a cost of 1.63 bps represents less than 10\% of the daily gross alpha.

While our primary analysis assumes execution at the market open, we also consider the liquidity dynamics at the market close as a robustness check.
The closing auction is a critical liquidity event which often commands higher volumes than the open.
Analysis of the closing bid-ask spreads for our investment universe reveals a similar, if not more favourable, liquidity profile.
Historically, spreads for Russell 1000 constituents tend to compress toward the end of the trading session.
For the Top-20 portfolios generated by the AI, closing spreads average approximately 1.50 bps across all signal horizons, slightly tighter than the opening spreads observed in Figure~\ref{fig:trading_costs}.
This confirms that the strategy's capacity is not constrained by the specific mechanics of the opening auction.

\end{document}